# Two Truths and A Lie? Benchmarking off-the-shelf LLMs for Requirements Quality Assessment – Performance, False Alarms, and Misses

Jannatul Shefa[1], Alejandro Salado[2], Paul Wach[2], and Taylan G. Topcu[1]

[1]Grado Department of Industrial and Systems Engineering, Virginia Tech, Blacksburg, Virginia, USA

[2]Department of Systems and Industrial Engineering, University of Arizona, Tucson, Arizona, USA

**Correspondence:** Taylan G. Topcu, PhD (ttopcu@vt.edu)

**Abstract**

Requirements engineering (RE) governs the quality of everything downstream in systems engineering (SE); defective requirements that survive review cycles propagate into design rework, schedule delays, and cost overruns. Because requirements are often written in natural language, recent advances in generative AI have raised expectations that large language models (LLMs) can absorb requirement quality assessment, a task otherwise slow and human expertise-intensive. Yet empirical evidence on whether LLMs can be trusted to do so remains scarce. This study presents the first benchmarking analysis of off-the-shelf LLM performance for requirement quality evaluation. Against an expert-derived ground truth built on INCOSE quality criteria, we evaluate ten models spanning two families (OpenAI and Anthropic) and five generations each, across one hundred independent runs, two requirement sets, and five sampling temperatures. Four contributions follow. First, we quantify a strongly asymmetric error profile: across all models and runs, the best-performing Anthropic model detects a median of only 47% of expert-identified issues while false-flagging 11%. Second, performance degrades significantly where SE judgment is required, as necessity and correctness issues are almost always missed. Third, generational progress is non-monotonic, so newer models cannot be assumed better. Fourth, this error behavior shifts only modestly and non-monotonically across sampling temperatures, indicating characteristic model deficiencies rather than inherent stochasticity. Off-the-shelf LLMs are therefore not yet trustworthy autonomous evaluators. Findings also warrant caution for Agentic AI developers: orchestrating these LLM modules in specialized architectures risks compounding these deficiencies rather than correcting them. Their defensible near-term role is human-in-the-loop decision support.

## 1. Introduction

Requirements in Systems Engineering (SE) play an integral role as they define the technical problem to be solved in a solution agnostic manner (Bahill & Madni, 2017; INCOSE, 2023a); thus, translate the ill-defined problem into a well-defined one that is amenable to solving effort (Simon, 1973). Since complex system development is a sequential process with tightly coupled task interdependencies (Simon, 1995; Hazelrigg, 1998), the elicitation of requirements are also critical as they constrain the trajectory of downstream activities by locking in resources and shaping the development effort (Klein et al., 2003; Doyle & Csete, 2011). Consequently, requirements engineering (RE) emerged as a subfield of SE to deal with the elicitation, analysis, and management of requirements through the system lifecycle. As such, numerous books have been published (Pohl, 1996; Nuseibeh & Easterbrook, 2000; Macaulay, 2012) and several international standards have been formalized for RE (IEEE, 2008, 2018).

Despite this emphasis on RE, many of the SE challenges continue to originate from requirements-related problems (Aurum & Wohlin, 2005). For instance, incomplete or missing information in requirement sets is often not discovered until much later in the lifecycle (Jayatilleke & Lai, 2018; Morkos et al., 2012). Additionally, requirement errors grow significantly more expensive to correct the later they are discovered. Compared to a baseline fix during the requirements phase, correction costs escalate to roughly ~8× in design, ~20× at manufacturing, ~100× at integration and test, and to over ~1500× once the system is in operation (Stecklein et al., 2004; Boehm et al., 2008). Requirement correction efforts are documented to be a leading cause of SE program delays and cost overruns (Boehm & Papaccio, 1988; Peña & Valerdi, 2015; Topcu & Shefa, 2025).

Given these challenges, advancing artificial intelligence (AI) capabilities and their use for SE (AI4SE) emerge as a promising avenue (McDermott et al., 2020; Allison et al., 2022). AI4SE research expanded rapidly in recent years, ranging from formulation of knowledge databases (Hirtz et al., 2002; Kitamura et al., 2004), virtual assistants (Bang et al., 2018; i Martin & Selva, 2019; Islas-Cota et al., 2022), and design evaluators (Fu et al., 1997; Guerlain et al., 1999). Significant research is being conducted on both human-AI dyads (Singh & Szajnfarber, 2025) and larger teams of humans with AI advisors (Gyory et al., 2021; Song et al., 2022a). Findings suggest that AI assistance can improve design performance, particularly if collaboration continues for extended periods of time (Viros I Martin & Selva, 2022). Studies investigating AI-assistance during design decision-making and found that designer self-confidence plays a critical role in their trust in AI, error attribution, and eventual adoption of AI guidelines (Chong et al., 2022). Current research focuses on agentic-AI approaches (Jiang et al., 2025; Massoudi & Fuge, 2025), with the notion that specialized agents, some of which could be LLM-based, could be orchestrated for improved performance. Although these are promising advances, effectiveness of AI in assisting SE tasks remains mostly an open research avenue (Topcu et al., 2025; Zhang et al., 2021).

Despite this growing body of research, evidence regarding the ability of LLMs to judge the quality of engineering requirements remains limited. Existing studies have primarily focused on feasibility and appearance of AI-assisted RE tasks, with little attention devoted to characterizing performance, error behavior, and evaluation reliability (Habib et al., 2025; Martins & Gorschek, 2017). Without a clearer understanding of these factors, the suitability of LLMs for trustworthy requirement quality evaluation remains uncertain. These challenges are further exacerbated by the fact that LLMs are stochastic tools that yield varying results (Su, 2025). This is especially important because requirement quality evaluation is not merely a language-processing task; it also requires contextual understanding, interdisciplinary SE knowledge (Vincenti, 1990), and judgment regarding how quality criteria should apply (Hull et al., 2010; INCOSE, 2023a). To that end, this study conducts a systematic evaluation of off-the-shelf LLMs for requirement quality assessment, pursuing three interrelated research questions (RQs). The first is:

***RQ1:*** *How well does an off-the-shelf LLM perform in terms of evaluation of requirements quality?*

We answer this question by comparing the responses of a cutting edge off-the-shelf LLM against a human-expert generated ground truth, repeated over multiple independent trials. This yields the distributions for the standard performance and error metrics, including true positive rate, true negative rate, false positive rate, false negative rate, accuracy, precision, and F1 scores. Based on these distributions and recognizing not all quality criteria are equally important; we seek a deeper understanding of error characteristics by investigating which quality aspects LLMs perform better with RQ2:

***RQ2**: How do error rates vary for various INCOSE Requirements Engineering quality criteria?*

Addressing RQ2 helps establish for which quality criteria LLM errors are more prevalent while documenting categories where it exhibits better performance. Finally, there is an expectation within the SE community that newer versions of foundation models will eventually perform on par with or better than human experts, particularly given that even earlier LLMs exhibited expert-like performance in the fields of medicine (Mbakwe et al., 2023; Takagi et al., 2023) and law (Katz et al., 2024). Hence, we ask:

***RQ3:** How does off-the-shelf LLM performance for RE evolve over-generations?*

Taken together, the findings of this study demonstrate higher performance in avoiding false alarms than in correctly detecting existing quality issues. The results also reveal substantial variation in both across repeated LLM evaluation attempts and across quality criteria. These findings shed light on where LLM-based tools can meaningfully support RE, what their limitations are, and in which aspects they would need to be supplemented for more sophisticated SE use cases. Overall, the study contributes empirical grounding to ongoing discussions on the trustworthiness of AI, its role in SE, and the pressing need for a human-in-the-loop approach for RE.

The rest of this paper is organized as follows. Section 2 reviews related work on AI for RE, AI-assisted requirement quality assessment, and the research gaps motivating this study. Section 3 describes the research methodology. Section 4 presents the findings in relation to the RQs. Section 5 discusses the findings, key contributions, practical implications, and limitations. Section 6 concludes the paper by summarizing the key insights, limitations, and directions for future research.

## 2. Literature review

This section reviews prior research relevant to AI-assisted requirement quality assessment. It first introduces broader AI4RE literature in Section 2.1, then discusses work specifically focused on requirement quality assessment in Section 2.2. Finally, it identifies the gaps motivating the present study in Section 2.3.

### 2.1 AI for RE (AI4RE)

RE has been one of the earlier adopters of AI4SE research in part because requirements are traditionally expressed in natural human language (e.g., English), and the software engineering community - who are also deeply concerned with RE - led the development of natural language processing algorithms (NLPs). NLP aims to render natural human language palatable to computers through executing a combination of tasks such as information retrieval, text classification, and language generation (Weizenbaum, 1966; Jurafsky & Martin, 2008). While earlier NLP lacked contextual awareness (Hayes-Roth, 1985); their transition to learning-based methods enabled better prediction of word sequences (Baker, 1975; Jelinek et al., 1977). A key breakthrough in NLP performance was the use of deep learning methods (Elman, 1990) that enabled selective retaining information for longer periods of time (Hochreiter & Schmidhuber, 1997; Sutskever et al., 2014). Finally, the introduction of transformer models, such as large language models (LLMs) NLPs exhibited a significant performance leap (Ashish, 2017).

LLMs fundamentally differ from recurrent network-based NLP approaches as they rely on self-attention mechanisms, which allow them to process sequences of information in parallel, as opposed to the sequential approach of neural networks. However, combined with the wide-spread availability of text-based RE data, LLMs significantly accelerated the use of AI for RE. Today, AI4RE pursues a plethora of objectives, including defining specifications (Dalpiaz et al., 2018), classification and extraction of requirements types from other engineering documents (Hey et al., 2020; Sainani et al., 2020), establishing traceability (Guo et al., 2017; Lin et al., 2021), identification of user needs (Han & Moghaddam, 2021), and knowledge extraction (Berquand et al., 2021).

Recent review studies (Norheim et al., 2024; Sonbol et al., 2022; Zhao et al., 2022) summarize AI4RE initiatives into four categories: requirements generation, analysis, translation, and quality assessment. Generation research focuses on extracting requirement templates (Reinhartz-Berger & Kemelman, 2020), mining user reviews to identify requirements-relevant information (De Araújo & Marcacini, 2021), and developing conversational assistants to support elicitation and formulation activities (Arora et al., 2023). Requirements analysis work include classification of requirement types (Kurtanović & Maalej, 2017), information extraction, user-need identification, consistency checking (Fischbach et al., 2021), and traceability between requirements and downstream artifacts such as design documents and test cases (Rahimi et al., 2014). Translation studies mainly pursue conversion of requirements expressed in natural language into structured or formal representations that are machine processible (Cosler et al., 2023; Hahn et al., 2022). This direction is consistent with ongoing SysML v2 transition efforts (Bajaj et al., 2022; LaVoie et al., 2026), which seek to support more precise, structured, and machine-interpretable system models that can reduce ambiguity and minimize human-introduced errors in SE practice (Salado & Wach, 2019). Nevertheless, these studies prioritize functionality over quality assessment (Habib et al., 2025).

### 2.2 AI for Requirements Quality Assessment

Requirement quality is concerned with determining whether requirements satisfy some quality criteria, such as completeness, unambiguity, verifiability, feasibility (Hull et al., 2010; INCOSE, 2023a). AI based quality assessment research is unexpectedly nascent, yet it can be considered on two levels: individual requirements and sets.

Research on individual requirement quality focuses on detecting ambiguities, completeness, adherence to requirement writing guidelines, or syntactic structures. For example, unsupervised learning mechanisms are used to assess completeness (Luitel et al., 2023). Fazelnia et al. (2024) investigated Natural Language Inference (NLI) for requirement specification defects. This NLI-based approach achieved an overall F1-score above 80% for detecting requirement specification defects and outperformed ChatGPT, NoRBERT, prompt-based classification, and probabilistic term-indicator baselines (Fazelnia et al., 2024). Jafari et al. (2023) examined ambiguity in software requirements using transformer-based approaches to detect ambiguous pronouns and resolve their likely antecedents. Their findings show that specialized encoder-based quality assessment models outperformed GPT-3.5 and GPT-4 for ambiguity resolution (Jafari et al., 2023). Talha et al. (2025) proposed a semiautomated NLP approach using SpanBERT and named entity recognition to detect anaphoric, coordination, and missing-condition ambiguities in functional requirements. Izhar et al. (2025) proposed a hybrid ambiguity-detection framework that combines rule-based techniques with text vectorization and a Random Forest classifier to identify ambiguous software requirements. Their study evaluated requirements across multiple project domains and reported strong performance, further showing that targeted ML-based approaches can support individual requirement quality assessment, particularly for ambiguity detection (Izhar et al., 2025). Lubos et al. (2024) evaluated Llama 2 for assessing individual software requirements against ISO 29148 quality characteristics. Using expert-derived ground, they showed that the LLM could identify quality issues, provide plausible explanations, and suggest improved requirement statements. However, open questions remain regarding stochastic variability, error behavior, and cross-model performance.

In the case of the requirement sets, quality assessment studies focus on identifying missing requirements (i.e., incompleteness), redundancies, and conflicts. Malik et al. (2024), for example, used semantic similarity analysis to identify redundancies and conflicts among requirements (Malik et al., 2024). Mahbub et al., 2024 have explored the use of GPT-4 to detect ambiguities, conflicts, and inconsistencies in requirement sets, finding that precision can vary considerably across cases, from 41% to 61%. This variability indicates that off-the-shelf LLMs are not yet sufficiently dependable for unsupervised requirement quality assessment (Mahbub et al., 2024). Fantechi et al. (2023) conducted an early evaluation of ChatGPT for detecting inconsistencies in requirement sets. Their results showed that ChatGPT could provide useful support, particularly for smaller requirement sets, but it could not outperform expert review. Melo (2026) demonstrated that LLMs, when combined with clustering and semantic similarity measures, can support requirement quality improvement in a large SE project, particularly for ambiguity, understandability, and redundancy analysis. However, it primarily evaluates improvement across review iterations rather than systematically characterizing the reliability, stochastic variability and error behavior of the LLMs.

Collectively, these studies highlight that AI-based requirement quality assessment is an active and promising area of research. However, the literature also points to unresolved questions about how reliably shelf LLMs evaluate engineered-system requirements across established quality criteria. The following section builds on these observations to discuss the research gaps that motivate the present study.

### 2.3 Research Gap

Despite growing interest in AI for RE, existing research leaves several important gaps in understanding whether off-the-shelf LLMs can reliably evaluate requirement quality for engineered systems. First, much of the AI4RE literature is based on data from software requirements, rather than requirements for engineered systems that bring hardware and software together. Importantly, engineered systems are bounded not only by software logic but also by laws of physics, manufacturing and operational constraints, and high design rework costs (Törngren & Grogan, 2018; Hennig et al., 2021). As a result, requirement quality issues in engineered systems carry broader downstream consequences and may require deeper systems-level reasoning than is typically needed in software-only contexts.

Second, prior AI4RE work has not sufficiently examined LLM performance across established requirement quality criteria (Cheng et al., 2026). This challenge is further complicated by the lack of unified terminology and taxonomy used for requirement quality assessment. Existing studies use heterogeneous terms for overlapping or closely related quality concepts, making it difficult to compare findings across studies or transfer methods across contexts (Wolf et al., 2025).

Third, the current literature remains limited in terms of evaluation maturity, reproducibility, and trustworthiness. Existing studies are often exploratory or proof-of-concept in nature, with limited methodological rigor, fragmented benchmarking practices, and inconsistent evaluation metrics (Cheng et al., 2026; Habib et al., 2025; Zadenoori et al., 2025). Moreover, existing work rarely cover all requirement quality aspects (Wolf et al., 2025). This limits the usefulness for SE practice, where requirements must satisfy multiple quality expectations simultaneously (Hull et al., 2010; INCOSE, 2023b). These limitations also extend to LLM model comparison over generations. Although LLM capabilities are evolving rapidly (Yang et al., 2025), existing studies, to the best of our knowledge, does not provide empirical evidence on whether newer model generations consistently improve requirement quality evaluation.

Fourth, LLMs lack transparent decision-making, are susceptible to hallucinations, exhibit limitations in numerical and technical reasoning, and possess uneven knowledge across specialized engineering domains (Kaddour et al., 2023; Liao & Vaughan, 2024; Zhang et al., 2025). On top of that LLMs are inherently stochastic tools (Ofsa & Topcu, 2025) whose outputs may vary across repeated executions, prompting conditions, and model versions. These characteristics render LLM-based requirement quality evaluation inherently error-prone (Fabbrini et al., 2001), specially manifested by false positive and false negative predictions (Norheim et al., 2024). Therefore, LLM performance and error behavior needs to be evaluated across multiple independent executions, with attention to the distribution and variability of their results (Cui & Alexander, 2026).

From a forward looking perspective, these gaps in knowledge pose a pressing need for research, particularly given emerging agentic AI approaches for RE and SE. Agentic AI methods often decompose complex tasks across multiple LLM-based agents or tool-specific modules, where each agent performs a specialized subtask (Hu et al., 2025; Sapkota et al., 2025). In such systems, an LLM-based requirement quality evaluator could function as a subsystem of a larger AI-assisted workflow. Therefore, understanding the reliability, variability, and error behavior of individual LLMs is a necessary form of component-level validation, or unit testing, before they are embedded in larger agentic systems at scale.

Taken together, these gaps emphasize the need for empirical evidence on how off-the-shelf LLMs behave when used as requirement quality evaluators for engineered systems. To address these gaps, this study provides the first benchmark of off-the-shelf LLM performance for requirement quality assessment. This benchmark is used to characterize overall evaluation performance, criterion-level error behavior, model-generation effects, and robustness across repeated evaluations. Next, we discuss the methodology.

## 3. Methodology

This study adopts a five-step methodology, as illustrated in Figure 1, to systematically generate, evaluate, and analyze AI-based assessments of requirement quality. The approach begins with the adaptation of two requirement sets derived from two different SE problems. Next, ground truth solutions are established for each sample requirement set through expert evaluations. Third, an experimental setup comprises two primary experimental conditions addressing the research questions, complemented by two sensitivity analyses that examine the robustness of the findings across sampling temperatures and requirement-set characteristics. Next, an automated workflow is employed to execute each experimental configuration and collect LLM evaluations across multiple independent runs. Finally, a quantitative analysis is conducted by comparing AI-generated evaluations against the expert-derived ground truth. Each step of the methodology is described in detail in the following subsections.

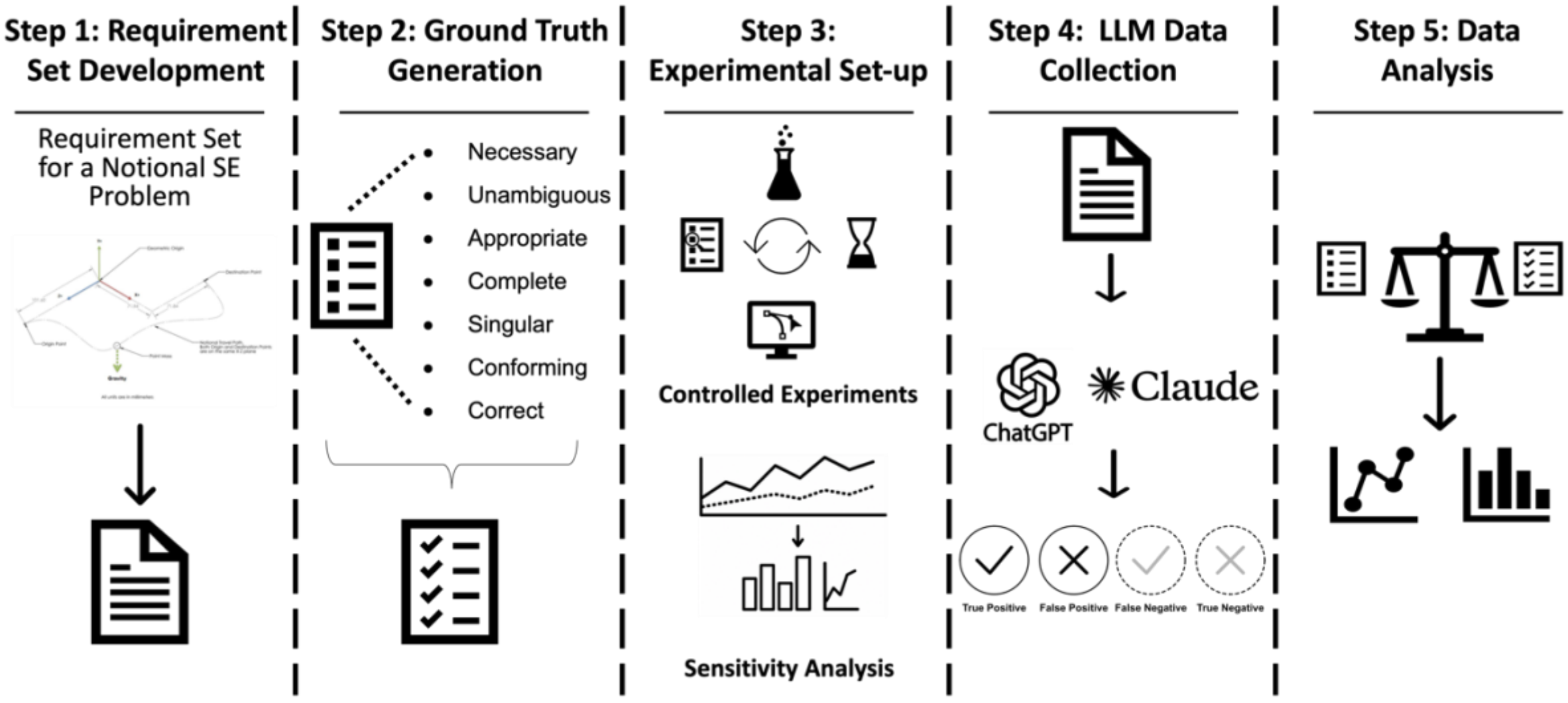

**Figure 1.** Overview of the research methodology.

### 3.1 Step 1: Requirement Set Development

Two requirement sets were used in this study, each representing a distinct source of human-generated requirements for a notional engineered system. The requirement sets are described below.

Requirement Set A was derived from a set of requirements for the Simple Deployment Mechanism (SDM) challenge, which was part of the "Astrobee Challenge Series" sponsored by the National Aeronautics and Space Administration (NASA) (Szajnfarber et al., 2020, 2023). The system of interest in this dataset is a robotic point-mass manipulator to serve on the International Space Station, which is required to move and return a point mass between defined positions based on a problem context and the Concept of Operations (ConOps) figure. From the SDM requirement set, 13 requirements were randomly selected to facilitate ground truth generation. This set was selected for its relatively balanced quality and well-defined system context. Importantly, independent developers were able successfully produce solutions from this requirement set that were subsequently evaluated for completeness and quality; and verified by independent NASA experts. This provides greater confidence that the requirements set provide a sufficiently realistic and representative basis for requirement quality evaluation challenges encountered in real-world SE practice. Requirement Set A is used for all baseline analysis in this study.

Requirement Set B was drawn from a study conducted by researchers at the University of Texas at Dallas (UT Dallas). In this study, UT Dallas researchers gave a guest lecture on requirements to 1,208 pre-service engineers with no prior SE education and provided them with a notional SE problem on a book retrieval system for people in wheelchairs. Participants were provided with a problem context description and a brief ConOps and were then instructed to work in various team sizes to generate as many requirements as possible within 20 minutes. The objective of the study was to understand the impact of team size on requirements variety (Nonso-Anyakwo & Summers, 2025b) and quantity (Nonso-Anyakwo & Summers, 2025a). This study produced a large pool of human-generated requirements characterized by a rich diversity of quality issues stemming from the participants' lack of SE training and the constrained problem formulation time. From this pool, 20 requirements were randomly selected to facilitate ground truth generation. Given that it is significantly less mature than Requirement Set A and its efficacy for leading to verifiable engineering solutions were never documented, Requirement Set B is used for sensitivity analysis in this study; and helps to assess whether the observed LLM performance patterns generalize for a substantially poorer requirement set.

Table 1 summarizes the key characteristics of both requirement sets across dimensions relevant to this study. Although both sets represent human-generated requirements for notional engineered systems, they differ substantially in authorship experience, requirement quality, requirement set size, and contextual richness. These contrasting profiles were an intentional choice to investigate whether the performance patterns observed in the requirement set A generalize to a more challenging scenario. Both Sets A and B were then categorized following Salado's work on Max-Neef's existential categories that enable a minimum set of categories to define a set of system requirements (Salado & Nilchiani, 2014). This helped organize the requirements in each set into functional, performance, environmental, and resource requirements.

**Table 1.** Requirement set comparison

| Dimension | Set A: NASA Astrobee SDM | Set B: UT Dallas Book Grabber |
|---|---|---|
| Number of requirements | 13 | 20 |
| Author experience level | Higher | Lower |
| Quality | Higher | Lower |
| Problem complexity | Higher | Lower |
| Authorship conditions | Unconstrained | 20-min limit |
| Requirement set origin | Crowdsourcing design competition | Classroom exercise |
| Role in this study | For baseline analysis | For sensitivity analysis |

### 3.2 Step 2: Ground Truth Generation

To create a basis of comparison, the research team evaluated each of the requirements in both requirement sets based on the nine International Council on Systems Engineering (INCOSE) requirements quality criteria summarized in Table 2 (INCOSE, 2023a). These criteria are widely adopted in the SE community as the standard for assessing the quality of requirements. For this phase of the study, the research team only focused on evaluating individual requirements as a first step towards requirement evaluation capability development using AI. Although the INCOSE criteria also include "Feasibility" and "Verifiability" for the individual requirement quality evaluation, the available requirement context did not provide sufficient information regarding organizational capabilities to reliably assess these aspects. Therefore, they were excluded from this analysis. In addition to the criteria mentioned in Table 2, No Issues and Unsure of Category were used to indicate cases with no identified issues and cases in which evaluators were unable to confidently map an observed issue to one of the predefined criteria, respectively.

**Table 2.** Requirement Quality Evaluation Criteria from INCOSE Handbook

| Criterion | Definition |
|---|---|
| Necessary | The need requirement statement defines capability, characteristic, constraint, or quality factor needed or required to satisfy a lifecycle concept, need, source, or higher-level requirement. |
| Appropriate | The specific intent and amount of detail of the need or requirement statement is appropriate to the level (the level of abstraction, organization, or system architecture) of the entity to which it refers. |
| Unambiguous | Need and requirement statements must be stated such that their intent is clear and can be interpreted in only one way by all intended audiences. |

| Criterion | Definition |
|---|---|
| Complete | The need statement sufficiently describes the necessary capability, characteristic, constraint, conditions, or quality factor to meet the lifecycle concept or source from which it was transformed. The requirement statement sufficiently describes the necessary capability, characteristic, constraint, conditions, or quality factor to meet the need, source, or higher-level |
| Singular | The need or requirement statement should state a single capability, characteristic, constraint, or quality factor. |
| Correct | The need statement must be an accurate representation of the lifecycle concept or source from which it was transformed. The requirement statement must be an accurate representation of the need, source, or higher-level requirement from which it was transformed. |
| Conforming | Statements and expressions of individual needs and requirements should conform to an approved standard pattern and style guide or standard for writing and managing needs and requirements. |

The ground truth generation process involved three evaluators within the research team with varied levels of expertise in RE; including two seasoned evaluators with decades of experience. Each evaluator independently assessed all requirements in both sets against the quality criteria, recording a binary judgment for each requirement-criterion pair. To minimize bias and reduce intercoder variability (Miles, 1994), the evaluators completed their assessments separately. Next, they participated in a group discussion to compare evaluations, align their interpretations of the quality criteria, resolve disagreements and reach consensus on the evaluation. Through this process, the reviewers agreed on the final ground truth used in the analysis.

This process resulted in a single unified ground truth label for each cell in the evaluation matrix**,** either positive (issue present, marked as **'**x**'**) or negative (no issue, left empty). For Requirement Set A, this produced an evaluation matrix of 117 **(**13 requirements × 9 criteria**)** binary assessment cells. For Requirement Set B, the same process yielded a matrix of 180 (20 requirements × 9 criteria) binary assessment cells. The resulting ground truth matrices serve as the basis for comparing AI-generated evaluations across the conditions, as applicable.

## 3.3 Step 3: Experimental Set-up

Two primary experiments were defined to address the three RQs using Requirement Set A. In addition to that two sensitivity analyses were conducted to evaluate the robustness of the findings under requirement contexts and alternative execution settings. The experimental and sensitivity analysis configurations are described below.

### *3.3.1 Experiment 1: Single-Model Baseline Evaluation (RQ1 and RQ2*)

The first experiment establishes a baseline characterization of off-the-shelf LLM performance for requirement quality evaluation using the latest Anthropic LLM model available at the time of data collection, Claude Opus 4.7 (released on April 16, 2026). Both the latest Claude and GPT models were screened initially, and Claude Opus 4.7 was selected for its stronger performance compared to the latest GPT model. Claude Opus 4.7 does not support temperature tuning, and the model was therefore evaluated using its default sampling configuration. 100 independent runs were executed on Requirement Set A as this provided a means to observe LLM performance variability to a certain extent while remaining within computational cost constraints. This experiment provides the basis for RQ1 and RQ2 by quantifying overall error rates and their variation across INCOSE quality criteria.

*3.3.2 Experiment 2: Cross-Generational Model Comparison (RQ3)*

This experiment extends the evaluation condition 1 across nine additional LLM variants spanning two model families, Anthropic's Claude and OpenAI's GPT, to investigate whether newer model generations outperform earlier ones in requirement quality evaluation. Within each family, models were selected to represent a range of successive generational releases to enable a performance comparison across generations. The Claude variants evaluated are Claude Opus 4.1 (released August 2025), Claude Sonnet 4.5 (released September 2025), Claude Opus 4.5 (released November 2025), Claude Sonnet 4.6 (released February 2026), and Claude Opus 4.7 (released April 2026). The GPT variants evaluated are GPT-4o (released May 2024), GPT-5 (released August 2025), GPT-5.1 (released November 2025), GPT-5.4 (released March 2026), and GPT-5.5 (released April 2026). Each model was evaluated using Requirement Set A with 100 independent runs per model, with all other pipeline configurations held constant. A sampling temperature of 0.7 was specified for models that supported explicit temperature configuration, while models that did not support temperature configuration were evaluated using their default settings.

*3.3.3 Sensitivity Probe 1: Requirement Set Robustness*

This sensitivity probe assesses whether the performance patterns observed across LLM generations for the requirement set A holds true for a substantially different requirement set. Consequently, this sensitivity analysis was conducted using Requirement Set B with all other pipeline configurations, described in 3.4.2, held constant.

*3.3.4 Sensitivity Probe 2: Temperature Robustness*

This probe investigates the sensitivity of LLM-based requirement quality evaluation to sampling temperature. This sensitivity probe was applied to Requirement Set A. GPT-5.4 was selected for this probe because it was the latest OpenAI/Anthropic model available at the time of data collection that supported temperature tuning while allowing other LLM model parameters to remain at their default settings. Five temperature settings were evaluated: 0.0, 0.3, 0.5, 0.7, and 1.0, spanning the range from deterministic to highly stochastic sampling behavior. For each temperature setting, 100 independent runs were executed under otherwise identical pipeline configurations. Table 3 summarizes the experimental Configuration, the associated RQs, the models and requirements set involved, and the temperature settings and number of runs applied.

**Table 3.** Summary of Experimental Set-up

| Configuration | Purpose | Model(s) | Requirement set | Temperature | # of Runs |
|---|---|---|---|---|---|
| Experiment 1 | RQ1, RQ2 | Claude Opus 4.7 | Set A | N/A | 100 |
| Experiment 2 | RQ3 | 10 Claude & GPT variants | Sets A | 0.7 Where Supported; Default Otherwise | 100 Per Model |
| Sensitivity Probe 1 | Sensitivity to Requirement Set | 10 Claude & GPT variants | Set B | 0.7 Where Supported; Default Otherwise | 100 Per Model |

| Sensitivity Probe 2 | Sensitivity to Temperature | GPT 5.4 | Set A | 0.0, 0.3, 0.5, 0.7, 1.0 | 100 Per Temperature |
|---|---|---|---|---|---|

### 3.4 Step 4: LLM Data Collection

The experimental configurations described in the previous section were implemented using an automated LLM evaluation workflow developed in n8n; with the objective of obtaining independent identical trials that can be analyzed in aggregate. n8n is a workflow automation platform to schedule repeated LLM evaluations, parse model outputs, and store structured results. Figure 2 illustrates the pipeline architecture. The pipeline operates as follows. A time-based trigger initiates each evaluation run at regular intervals. An execution number is assigned to each run for traceability. Two Google Sheets are created per run: one for the structured evaluation matrix output and one for capturing raw LLM responses. The requirement list, problem description, and ConOps are retrieved from a source sheet, formatted into an LLM-readable structure by the Process Sheet Data node, and passed to the LLM via the Evaluate Requirements node. The evaluation used a zero-shot prompting approach (Kojima et al., 2022), where the prompt provided the requirement quality categories, brief explanations of what each category meant, and instructions for evaluating each requirement against those categories. Although prior work shows that prompting strategy can substantially influence LLM performance (Topcu et al., 2025), the objective of the current study is to establish a baseline characterization of off-the-shelf LLM behavior for requirement quality evaluation. Hence, the zero-shot prompting approach was intentionally selected for this purpose without task-specific finetuning, example-based conditioning, domain-specific retrieval augmentation, or other model customization. The prompt used in this study is provided in Appendix A. The LLM response is parsed by the Parse AI Output node, which extracts the structured JSON output. The Transform to Matrix node converts the parsed output into the binary evaluation matrix format. The binary evaluation matrix is appended to the Google Sheet. Additionally, the LLM-generated raw response is appended to a separate sheet for archival.

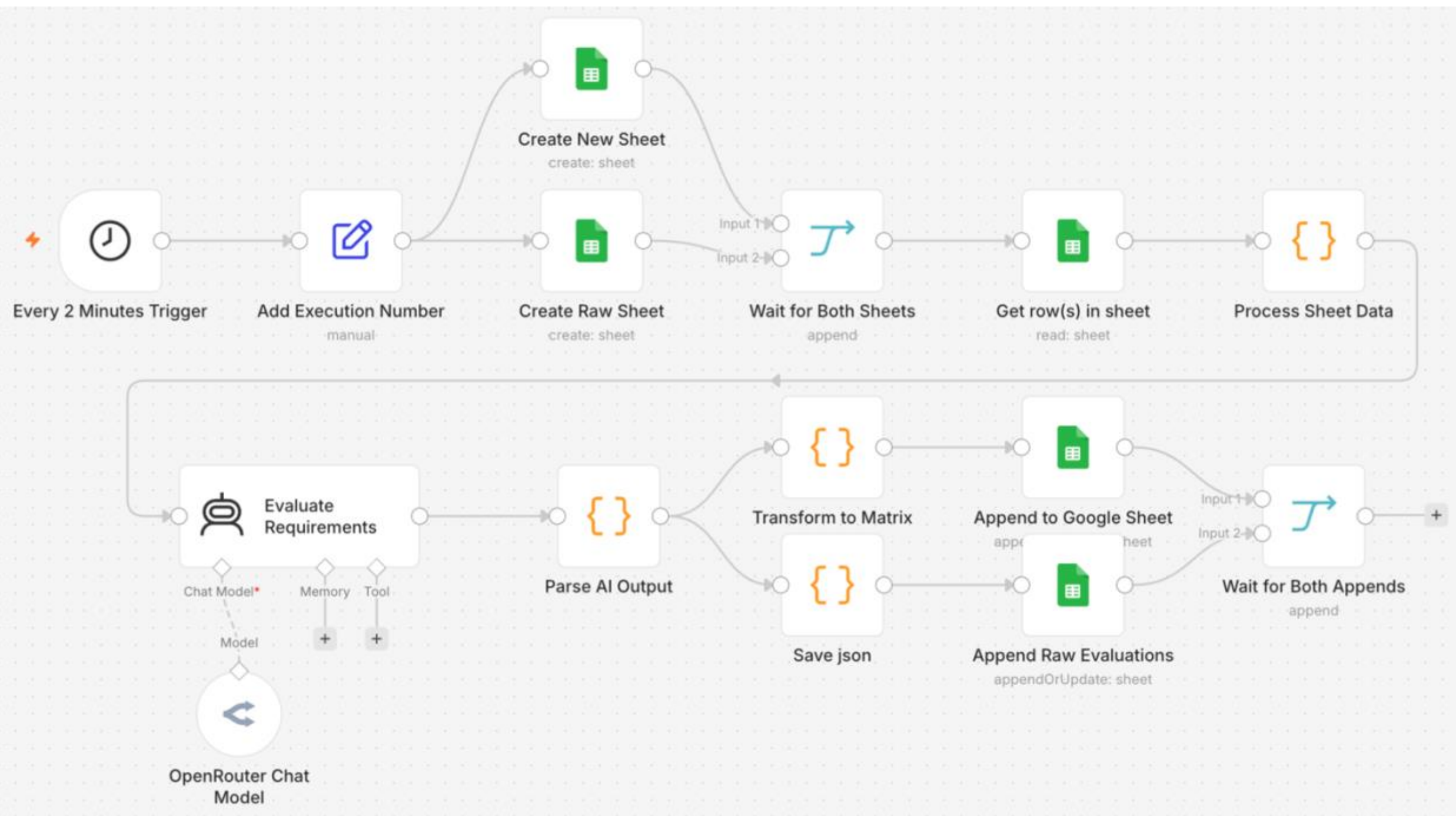

**Figure 2.** Automated LLM evaluation pipeline implemented in n8n.

LLMs used for analysis are accessed via the OpenRouter API. Within the automated workflow, the AI was prompted to evaluate if a problem exists for each of the requirements by using its internal knowledge as well as the problem-relevant information (ConOps and Project Description). The prompt instructed the model to evaluate each requirement individually against all nine quality criteria and return a structured JSON response containing, for each criterion violation, the criterion ID, criterion name, and a brief explanation. The output format was specified to facilitate automated parsing. The LLM outputs generated through the automated workflow for all experimental conditions served as the basis for all subsequent data analyses as described in Section 3.5.

**3.5 Step 5: Data Analysis**

The following procedure was followed to analyze LLM performance for requirements evaluation. First, for each run, AI-generated outcomes were compared against the expert-generated ground truth. Each comparison resulted in one of four possible outcomes: correct detection, correct non-detection, false detection, or missed detection. Next, building on this pairwise comparison, each AI decision was classified using standard confusion matrix logic (Foody, 2002; Heydarian et al., 2022) as follows:

- True Positive (TP): AI indicates a quality issue that is also captured in the ground truth.
- True Negative (TN): AI does not indicate an issue that does not exist in the ground truth.
- False Positive (FP): AI indicates a quality issue when the ground truth indicates no issue. This is also known as a Type I error or a false alarm (Robinson et al., 1998).
- False Negative (FN): AI does not indicate a quality issue when the ground truth indicates an issue. This is also known as a Type II error or a miss (Robinson et al., 1998).

Finally, to enable fair comparison across quality criteria, we normalized performance rates rather than reporting raw counts. Seven normalized metrics were computed as follows:

i. True positive rate (TPR) /Recall: Among all real issues present, what fraction did the LLM correctly identify? Higher TPR is better.

$$TPR = \frac{TP}{Ground\ Truth\ Total\ Positve}$$

ii. True negative rate (TNR): Among all non-issues, what fraction were correctly left unflagged? Higher TNR is better.

$$TNR = \frac{TN}{Ground\ Truth\ Total\ Negative}$$

iii. False positive rate (FPR): Among all non-issues, what fraction were incorrectly flagged? Lower FPR is better.

$$FPR = \frac{FP}{Ground\ Truth\ Total\ Negative}$$

iv. False negative rate (FNR): Among all real issues present, what fraction were missed? Lower FNR is better.

$$FNR = \frac{FN}{Ground\ Truth\ Total\ Positve}$$

v. Accuracy: Out of all predictions made, what fraction were correct? Higher Accuracy is better.

$$Accuracy = \frac{TP + TN}{TP + TN + FP + FN}$$

vi. Precision: Among all issues flagged by the LLM, what fraction were actual issues? Higher Precision is better.

$$Precision = \frac{TP}{TP + FP}$$

vii. F1 Score: Harmonic mean of Precision and TPR balancing both in a single metric. Higher F1 is better.

$$F1 = \frac{2 \times Precision \times TPR}{Precision + TPR}$$

Here, TPR reflects correct issue detection, and TNR reflects correct non-issue identification (Luque et al., 2019). FPR and FNR capture complementary error rates associated with incorrect issue flagging and missed issue detection, also known as Type I and Type II errors, respectively (Cuellar, 2025). Accuracy reflects the overall proportion of correctly classified instances across both issue and non-issue categories. However, both ground truth matrices are imbalanced. Under such conditions, accuracy can be misleading. For example, quality violations were observed in approximately 26% of all cells of requirement set A, the ground truth. As a result, a model that flags no issues would still achieve 74% accuracy by consistently predicting the majority class. Therefore, precision and F1-score become the primary metrics of interest, as they penalize both false positives and false negatives and provide a more reliable assessment under class imbalance.

The performance of each experimental condition described in Step 3 was assessed by comparing AI-generated evaluation outcomes against the expert-generated ground truth. For each run, the seven metrics were computed. Next, each metric was examined using boxplots to characterize the central tendency, dispersion, and outliers of each distribution across the 100 runs for each condition.

## 4. Findings

This section presents the findings of the study around the three research questions. Section 4.1 addresses RQ1 by characterizing the overall performance and error rates of an off-the-shelf LLM in evaluating requirement quality. Section 4.2 addresses RQ2 by examining how detection performance and error rates vary across individual INCOSE requirement quality criteria. Section 4.3 addresses RQ3 by comparing performance across ten LLM variants from two model families to assess whether generational advancement produces consistent improvement. Section 4.4 presents findings from a sensitivity analysis focusing on a poorer quality requirement set and varying sampling temperature to gauge the generalizability of findings. It is important to note that the variability observed across runs reflects the inherently probabilistic nature of LLM inference and should be treated as a characteristic of LLM behavior rather than experimental error. We discuss each of the subsections below.

### 4.1 RQ 1: How well does an off-the-shelf LLM perform in terms of evaluation of requirements quality?

Figure 3 presents the distribution of the detection metrics across 100 executions for requirement quality evaluation using Claude Opus 4.7 presented as boxplots, providing a comprehensive view of the performance of an off-the-shelf LLM in identifying requirement quality issues. The results reveal a pronounced imbalance between the LLM's ability to identify the existing requirement quality issues for a complex engineered system problem taken from a NASA crowdsourcing challenge.

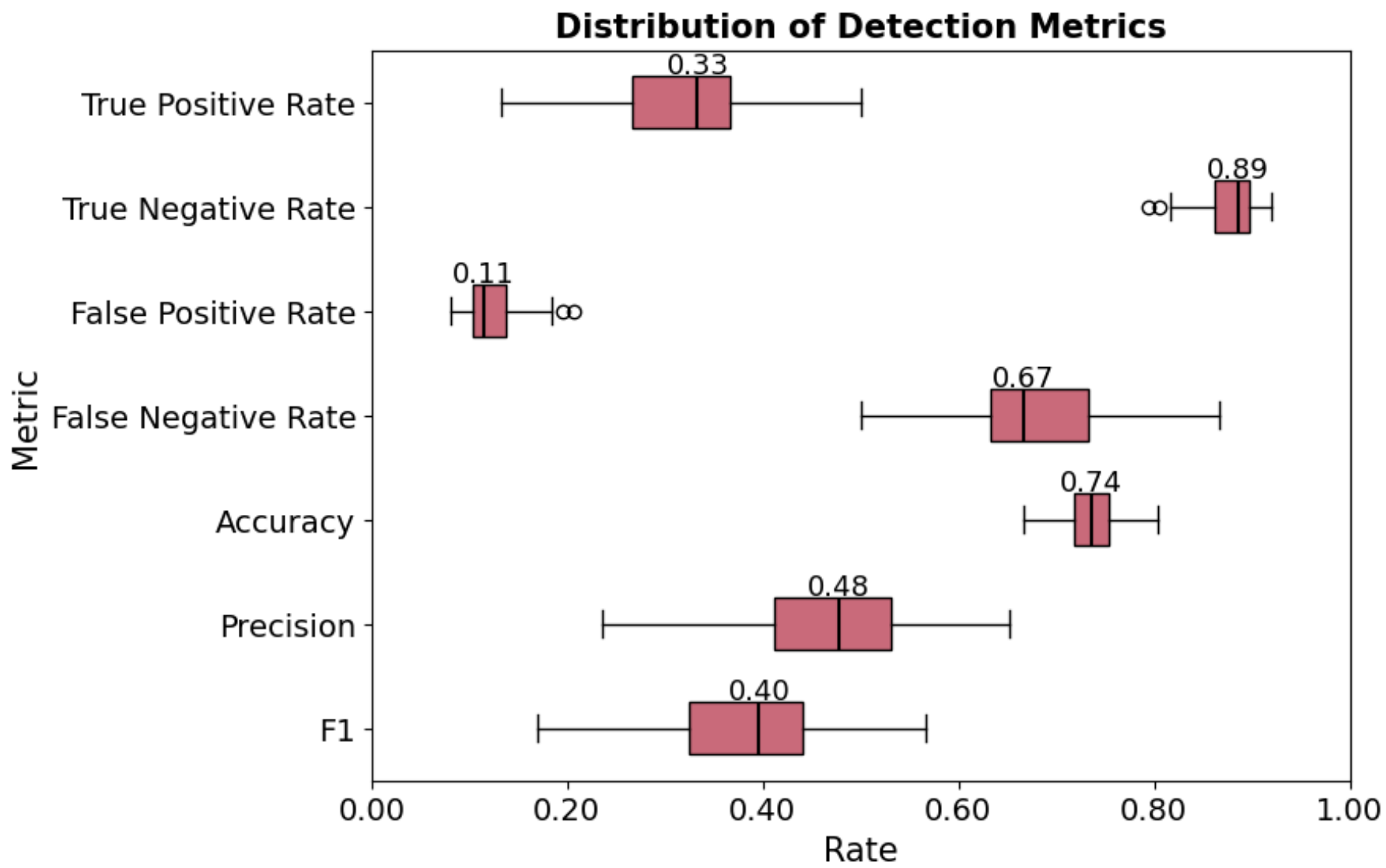

**Figure 3.** Distribution of Requirement Quality Issue Detection Rates for Claude Opus 4.7

The median TPR is relatively low at 0.33, indicating that the LLM correctly identifies only 33% of existing requirement quality issues. This behavior is directly reflected in the substantially higher FNR (i.e., Type II Error Rate) median of 0.67, suggesting that nearly two-thirds of actual quality issues are consistently missed. In terms of distribution characteristics, both TPR and FNR exhibit broad interquartile ranges further suggesting substantial variability high variability in requirements quality assessment performance. TPR values range approximately from 0.13 to 0.50, while FNR ranges from roughly 0.50 to 0.87. This indicates that even under the most favorable execution conditions, the model detects only half of the requirement quality issues. Further, the comparatively large standard deviations of 0.08 for both TPR and FNR indicate instability and inconsistency in issue-detection performance across executions. The TPR distribution exhibits a left skew, indicating that while most executions cluster around lower detection performance, a limited number of executions experience even poorer issue-detection capability. Conversely, a corresponding right skew in FNR distribution suggests that although most executions already demonstrate higher missed-detection rates, some executions exhibit substantially severe missed-detection behavior.

In contrast, Opus 4.7 demonstrates substantially stronger performance in avoiding false alarms. The median TNR is high at 0.89, indicating that the LLM correctly leaves the majority of non-issues unflagged. Similarly, the median FPR (Type I Error) is comparatively low at 0.11, suggesting that only 11% false-flag claims or in other words, non-issues incorrectly classified as problematic. Both TNR and FPR exhibit narrower distributions compared to TPR and FNR, with TNR ranging approximately between 0.79 and 0.92 and FPR ranging between 0.08 and 0.21. That being said, although this false alarm rate appears relatively modest, it may still result in unnecessary rework and mistrust in AI. The compact interquartile ranges and lower standard deviations of 0.03 for both TNR and FPR indicate comparatively stable behavior across executions. The boxplots also reveal mild left skew for TNR and right skew for FPR, indicating that most executions consistently achieve high true negative performance and low false positive performance with occasional poorer-performing executions.

The aggregate performance metrics further reinforce these observations. The median Accuracy is 0.74, with values distributed approximately between 0.67 and 0.80, indicating that the model correctly classifies nearly three-quarters of evaluated requirements. However, this comparatively high accuracy should be interpreted cautiously because it is disproportionately influenced by the model's strong true negative performance (median TNR = 0.89) relative to its substantially weaker true positive performance (median TPR = 0.33). The tight spread, relatively low standard deviation of 0.03, and near-symmetrical behavior of the Accuracy distribution also suggest that independent executions achieve moderately consistent performance.

The Precision metric reveals a concerning issue. The median Precision of 0.48, with observed values ranging roughly from 0.24 to 0.65, indicates that almost half of the flagged instances do not actually contain quality issues. In the worst-performing cases, approximately 76% of flagged issues are incorrect classifications. This substantially reduces the reliability and trust in LLM-assisted requirement evaluation. The highest standard deviation of 0.10 among all the metrics further indicates considerable variability in the reliability of predicted issues across executions. The Precision distribution also demonstrates mild left skew, indicating that while most executions show already poor precision, a smaller number of executions demonstrate even poorer prediction reliability.

The median F1-score of 0.40, ranging approximately from 0.17 to 0.57, further highlights the model's struggle in balancing precision and TPR simultaneously. The relatively high standard deviation of 0.09 further suggests notable instability in maintaining this balance. The F1 distribution also shows a slight left skew, suggesting that although most executions cluster around a moderately better balance between precision and recall, occasional executions experience substantially degraded overall detection effectiveness. Finally, none of the evaluation metrics demonstrate near-perfect performance. This indicates that substantial detection errors persist even in the best-performing trials highlighting LLM fallibility.

Next, we examine the criterion-level distributions of the error rates to provide a more fine-grained analysis of the LLM's performance across quality criteria.

### 4.2 RQ 2: How do error rates vary for various INCOSE Requirements Engineering quality criteria?

This section presents the distributions of criterion-level error rates, i.e., FPR and FNR, for Claude Opus 4.7. While all seven metrics discussed in section 3.5 were computed for each INCOSE quality criterion, this section focuses on the error-rate metrics, as these directly address RQ2. The criterion-level distribution of the rest of the metrics are provided in Appendix Figures B1 – B5, respectively, to provide a complete view of the LLM's performance.

It is important to note that several criteria do not display boxplots for different metrics. This occurs when no ground truth instances exist for a given quality criteria for a given metric, making the corresponding metric mathematically undefined. For example, no instances of "No Issues" and "Unsure of Category" exist in the expert-derived ground truth dataset because all requirements violated at least one quality criterion, and each issue could be assigned to a specific quality criterion. As a result, the denominator required for computing FNR becomes zero, resulting in mathematically undefined values and no plotted box distribution of FNR is present for these two categories. In contrast, criteria whose boxplots collapse to a single line at 0.00 or 1.00 indicate consistent outcomes across executions. For instance, the collapsed FNR distribution for "Necessary" at 1.00 indicates that the model always failed to detect any Necessity-related issues across all executions.

#### *4.2.1 Type I Errors or False Flags (False Positive Rates) Per Quality Criterion*

Figure 4 presents the distribution of FPR across the quality criteria, revealing the tendency to raise false alarms in general, with substantial variation among categories. The highest false alarm tendency of LLM among all criteria was observed in the Conformance criterion. Detection of Conformance demonstrated an FPR spanning from approximately 0.00 to 0.67 with a median of 0.67 and a mean near 0.45. The substantial spread, large standard deviations of 0.25, and skewness patterns indicate highly inconsistent execution-level behavior, where the model frequently misclassifies valid requirements as problematic. The second most concerning among this set was regarding detection of ambiguities, one of the most practically significant dimensions of requirement quality assessment. FPR distribution for Ambiguity spans approximately 0.11 to 0.78 with a median of 0.33 and a mean of ~0.30. This was accompanied by a broad spread and summarized with a standard deviation of 0.20; indicating substantial variability in false-alarm behavior. The mild right skew in FPR suggest that while most executions yield relatively lower Type I error rates, some executions produce substantially elevated false alarm rates.

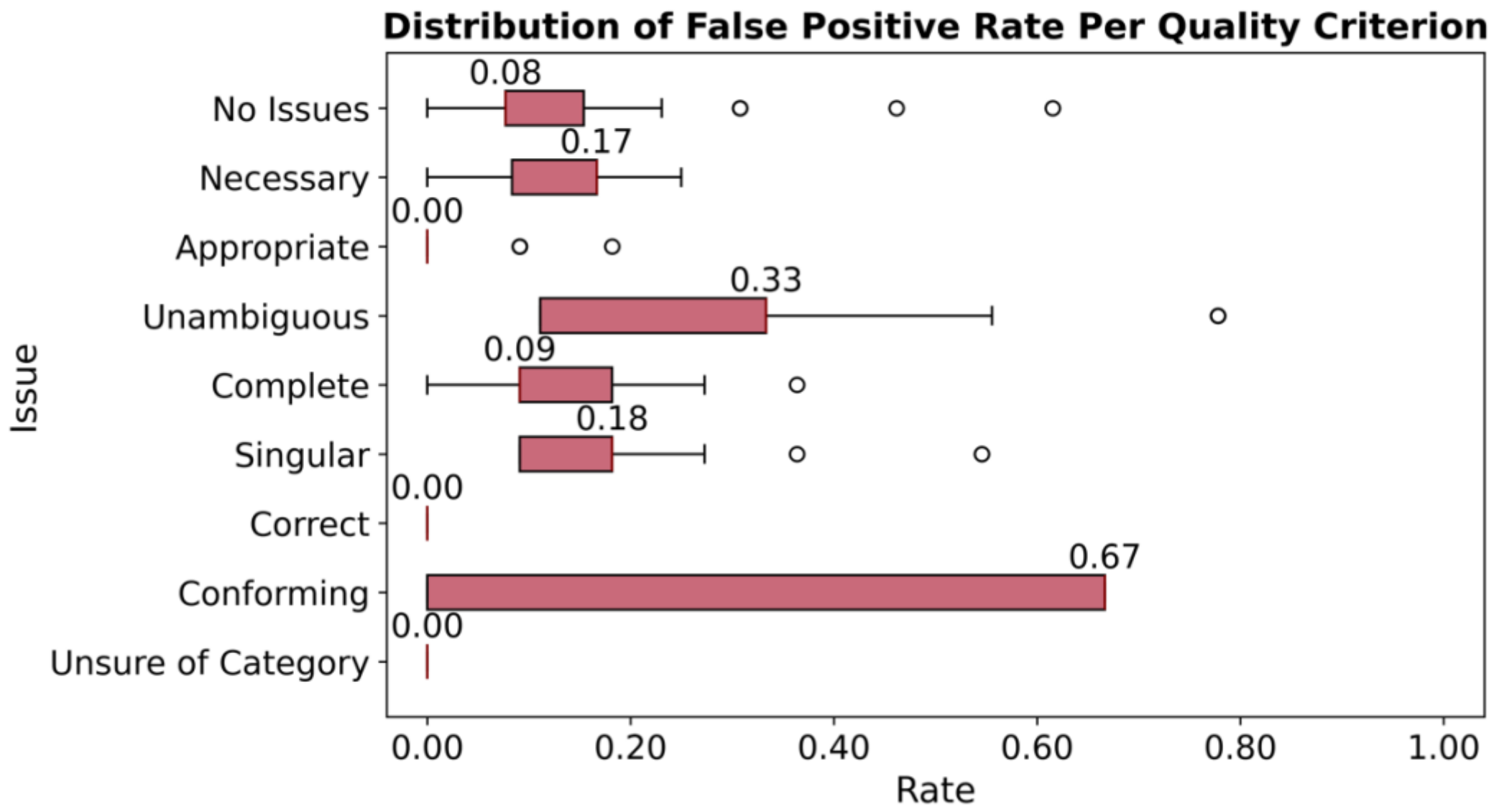

**Figure 4.** FPR (Type I Error) Distribution by INCOSE Requirement Quality Criterion

Relatively more moderate, yet still noteworthy, false flags behavior is observed regarding Completeness, Necessity, and Singularity criteria; arguably some of the most critical aspects of requirement quality assessment. Detection of Complete demonstrates a median FPR of 0.09 and a mean of 0.13. Detection of Necessary achieves a median FPR of 0.17 with a mean near 0.14. Detection of Singular exhibits similar behavior, with by a median FPR of 0.18 and a mean of 0.17. LLM performance distribution across all these three criteria have low standard deviations (≤ 0.09) for FPR, indicating reasonably stable false-alarm behavior across executions. Their relatively compact but skewed distributions indicate that the LLM consistently avoids excessive false alarms, although occasional executions still incorrectly flag valid requirements. the LLM exhibited exceptional capability in avoiding false alarms for Correctness and Appropriateness, each achieving collapsed FPR distributions at 0.00. This trend was also observed in declarations of Unsure of Category. These invariant distributions with low standard deviation (≤ 0.04) highlight a desirable level of consistency. Identification of No Issues also demonstrates a strong and stable performance, with a median FPR of 0.08 and a mean FPR of 0.11. The relatively small standard deviations of 0.10 further indicate comparatively stable execution-level behavior. FPR values cluster near zero with occasional elevated false-positive executions while the mild right skew in FPR distribution indicate that most executions successfully avoid false alarms in this criterion.

*4.2.2 Type II Errors or Misses (False Negative Rates) Per Quality Criterion*

Missing requirement quality issues is a critical failure mode as their implications are often undetected until much later in the design process, ultimately leading to design rework that causes cost and schedule overruns. Figure 5 presents the distribution of Type II errors or misses for each of the INCOSE requirement quality criterion; revealing significantly high miss rates with substantial variation across criteria. The following discussion is organized from the criteria exhibiting the highest missed detection performance to those demonstrating comparatively reliable performance.

The most concerning finding is the LLM's inability to detect issues related to Correctness and Necessity, arguably two of the most fundamental dimensions of requirement quality. Necessary represents the most extreme failure case. FNR remains identically 1.00 with zero variance and a mean of 1.00. These collapsed complementary distributions indicate that the model completely fails to detect any necessity-related issue throughout the entire evaluation process. Missed detection rate for Correct have a high median of 0.89 with a mean near 0.91 and a range of 0.89 to 1.00. The collapsed FNR distribution along with very small standard deviations of 0.03 indicate that LLM almost always fails to identify Correctness-related issues across executions. These findings raise a significant practical concern. Because the model consistently fails to detect Correctness- and Necessity-related issues, practitioners should exercise caution when using off the shelf LLM for these aspects of requirement quality assessment.

More encouraging, but still concerning, results are observed for detection trends of Appropriate, Unambiguous, Complete, Singular, and Conforming each exhibit median FNR values of 0.50. However, the shape, spread, and mean values of the FNR distributions differ considerably across categories. Type II error of Complete demonstrates the most extreme execution-level instability among all categories. FNR ranges from 0.00 to 1.00 with a median of 0.50 and a mean of 0.59, accompanied by a collapsed interquartile range and outliers near both extremes. The largest standard deviations of 0.39 further indicate extreme variability and lack of consistency across executions. These distributions indicate that the LLM is highly unreliable capability for identifying completeness-related defects.

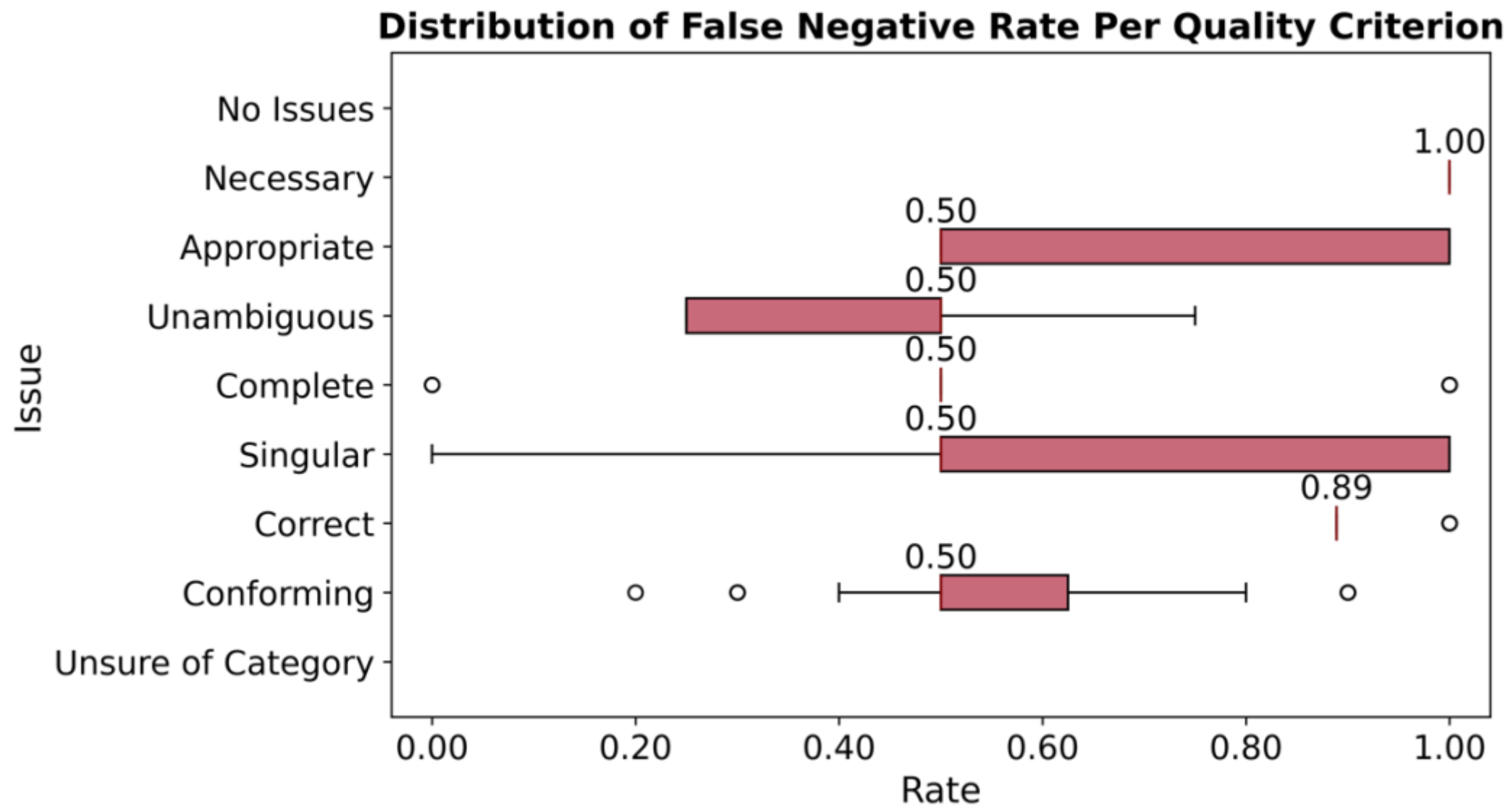

**Figure 5.** FNR (Type II Error) Distribution by INCOSE Requirement Quality Criterion

Similarly, substantial limitations are observed for the detection of Singularity- and Appropriateness-related issues. Missed detection trend for Appropriate presents a strongly asymmetric shape despite the balanced median of 0.50. The FNR distribution spans 0.50 to 1.00 and a mean of 0.71, and is correspondingly concentrated toward the upper end. The larger standard deviations of 0.24 for further reflect substantial variability across executions despite the balanced central tendency. These paired distributions suggest that the LLM struggles to reliably identify Appropriateness-related issues and frequently leaves such issues undetected. For the Singular criterion, FNR spans the full 0.00 – 1.00 range with median of 0.50 and a mean of 0.71. The left-skewed FNR distribution with elongated lower-tail behavior and very high standard deviations of 0.35, further indicates that the LLM's ability to detect singularity-related issues is highly poor and unpredictable across runs.

More encouraging results are observed for Conformance and Ambiguity, although missed detections remain common. Missed detection rate for Conforming exhibits moderately stable performance. FNR ranges from 0.20 to 0.90 with a median of 0.50 and a mean of 0.56. The moderate standard deviations of 0.16 indicate noticeable but comparatively lower variability. The mild left skew in FNR suggest that although most executions cluster around moderate issue-detection capability, some executions achieve improved detection performance accompanied by lower missed-detection rates. Unambiguous demonstrates the most consistent detection behavior among the all the criteria with a median FNR of 0.50 and a mean of 0.43. The comparatively lower standard deviations of 0.12 further indicate relatively stable execution-level behavior. The roughly symmetric and relatively narrow spread indicates that the LLM demonstrates neither strongly favoring detection nor consistently missing ambiguity-related issues.

Finally, No Issues and Unsure of Category produce no boxplot distributions in either metric because no true positive instances exist within the ground truth evaluation for these categories. Consequently, FNR values are mathematically undefined and therefore omitted from Figure 5.

### 4.3 RQ3: How does off-the-shelf LLM performance for RE evolve over-generations?

Figure 6 compares normalized performance metrics across model generations to evaluate whether newer LLM generations outperform earlier ones in requirement quality evaluation. The results reveal that performance improvements across generations exhibit fluctuating non-monotonically increasing behavior; suggesting that newer models may not lead to better performance. Additionally, if a newer model is performing better in a metric, it does not mean that it achieves stronger performance across all metrics (e.g., an increase in TPR performance may not mean less Type I errors). The distributions also vary substantially in spread, skewness, and stability across independent runs, indicating that newer generations do not consistently produce more reliable or predictable behavior.

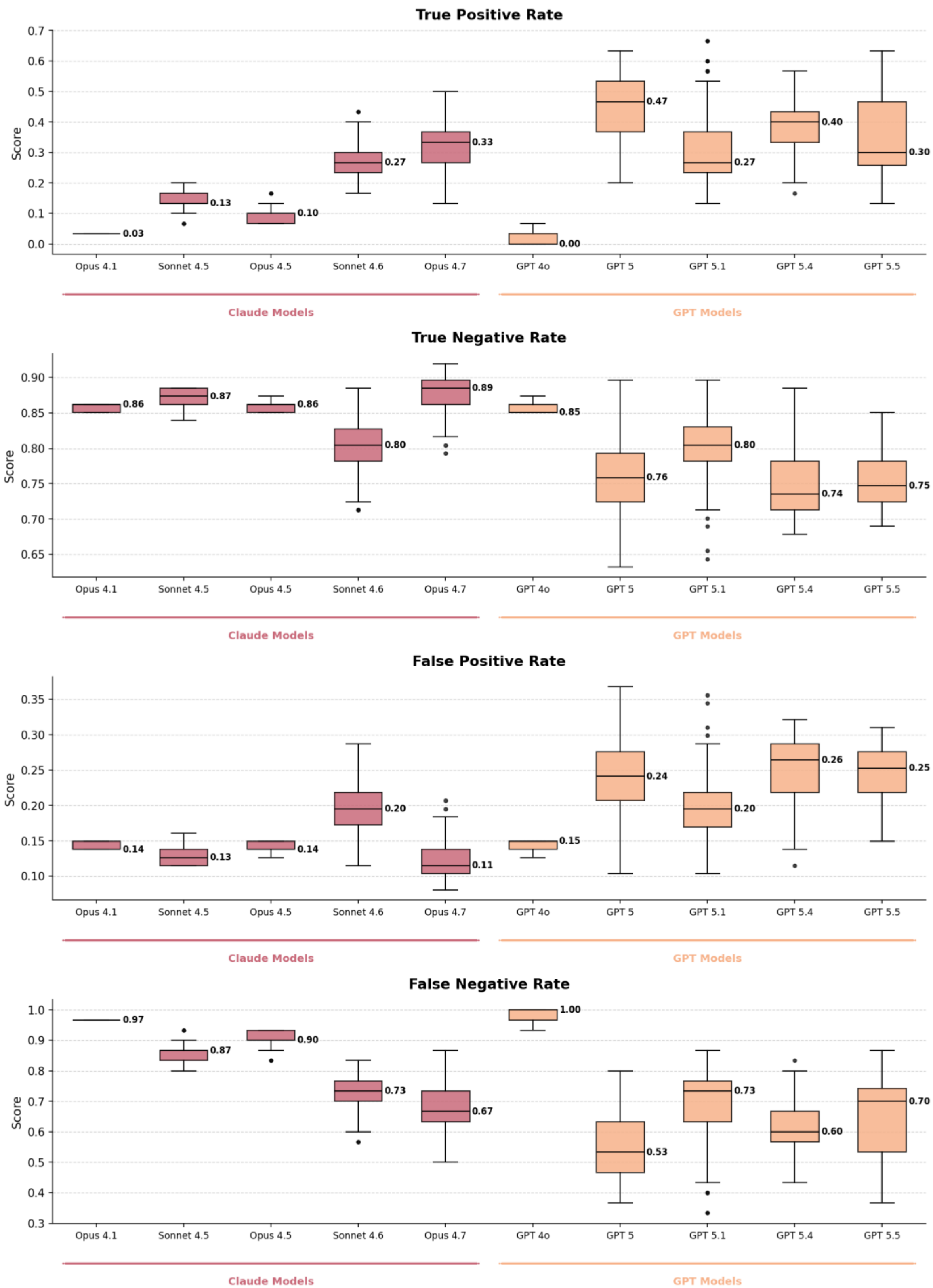

**Figure 6.** Normalized Performance Rates Across Generations and Models

In Figure 6, both the Claude and GPT families exhibit that improvement is inconsistent and resembles a fluctuating non-monotonic behavior. Within the Claude family, median TPR increases substantially from 0.03 for Claude Opus 4.1 to 0.27 for Claude Sonnet 4.6 and further to 0.33 for Claude Opus 4.7, accompanied by corresponding reductions in FNR from 0.97 to 0.73 and 0.67, respectively. However, the progression is not strictly linear. Claude Opus 4.5 achieves a lower median TPR of 0.10 than Claude Sonnet 4.5 at 0.13 despite being a later release variant. Similarly, within the GPT family, GPT-4o exhibits almost complete detection failure with a median TPR of 0.00 and FNR of 1.00, while GPT-5 improves substantially to a median TPR of 0.47 and FNR of 0.53. However, GPT-5.1 declines to a median TPR of 0.27, GPT-5.4 partially recovers to 0.40, and GPT-5.5 again decreases to 0.30.

The distributional characteristics further reinforce this instability. Earlier models, such as GPT-4o and Claude Opus 4.1, exhibit tight TPR and FNR distributions. In contrast, newer models generally exhibit wider spread and more pronounced skewness, reflecting increased variability in detection outcomes. For Claude models, the increase in variability is almost monotonic, with TPR ranges and standard deviations expanding progressively across generations: Claude Opus 4.1 (0.03–0.03, std=0.00), Claude Opus 4.5 (0.07–0.17, std=0.025), Claude Sonnet 4.5 (0.07–0.20, std=0.026), Claude Sonnet 4.6 (0.17–0.43, std=0.061), and Claude Opus 4.7 (0.13–0.50, std=0.075). Claude Opus 4.7 exhibits the widest distribution with the highest standard deviation among the Claude variants, reflecting both the broadest range and the most pronounced skewness. For GPT models, the increase in variability is more inconsistent, as reflected in the TPR ranges and standard deviations across generations: GPT-4o (0.00–0.07, std=0.017), GPT-5.1 (0.13–0.67, std=0.118), GPT-5.4 (0.17–0.57, std=0.086), GPT-5.5 (0.13–0.63, std=0.129), and GPT-5 (0.20–0.63, std=0.103). Notably, GPT-5.1 achieves the highest upper bound across all GPT variants at 0.67, yet GPT-5.5 exhibits the greatest standard deviation (0.129), underscoring that non-monotonic behavior and variability increases within the GPT family.

A similar non-monotonic trend appears in false alarm behavior. Claude models generally maintain stronger TNR and lower FPR than GPT models, although later Claude variants do not always improve consistently. Claude Opus 4.7 achieves the strongest non-issue classification performance among Claude models with a median TNR of 0.89 and FPR of 0.11, whereas Claude Sonnet 4.6 demonstrates noticeably weaker performance with TNR and FPR medians of 0.80 and 0.20, respectively. Within the GPT family, GPT-5 improves issue detection substantially relative to GPT-4o but simultaneously increases false alarms, producing a median FPR of 0.24 compared to 0.15 for GPT-4o. Later GPT variants also fail to demonstrate consistent progression. GPT-5.1 reduces FPR to 0.20, whereas GPT-5.4 increases it again to 0.26 despite improved TPR performance. GPT-5.5 similarly maintains elevated false positive behavior with an FPR median of 0.25.

The distributional characteristics further demonstrate this instability. GPT-4o and Claude Opus 4.1 both show near-identical TNR ranges of 0.85–0.87 with standard deviations of just 0.006, reflecting highly stable but rigid false alarm behavior. In contrast, newer GPT generations display broader distributions and stronger skewness patterns. GPT-5 exhibits the widest FPR spread among all models (0.10–0.37, std=0.050), followed closely by GPT-5.1 (0.10–0.36, std=0.047) and GPT-5.4 (0.12–0.32, std=0.045), which reflect an upward shift relative to earlier generations. Claude Opus 4.7, by contrast, demonstrates a comparatively more contained FPR distribution (0.08–0.21, std=0.024). These findings indicate that newer generations do not uniformly improve the reliability of false alarm avoidance behavior: while median TNR performance improves in some cases, the widening distributional spread, particularly pronounced in GPT-5, GPT-5.1, and GPT-5.4, points to increasing execution-level unpredictability in false-positive control. Thus, later generations often improve maximum achievable performance while simultaneously introducing greater variability and less predictable execution-level outcomes.

Aggregate metrics in Figure 7 further reveal the absence of a strictly progressive generational trend within either model family. For Claude models, Accuracy improves across generations, from a median of 0.65 for both Claude Opus 4.1 and Claude Opus 4.5, to 0.68 for Claude Sonnet 4.5, and reaching a peak of 0.74 for Claude Opus 4.7, suggesting a broadly upward trajectory, though not a uniform one.

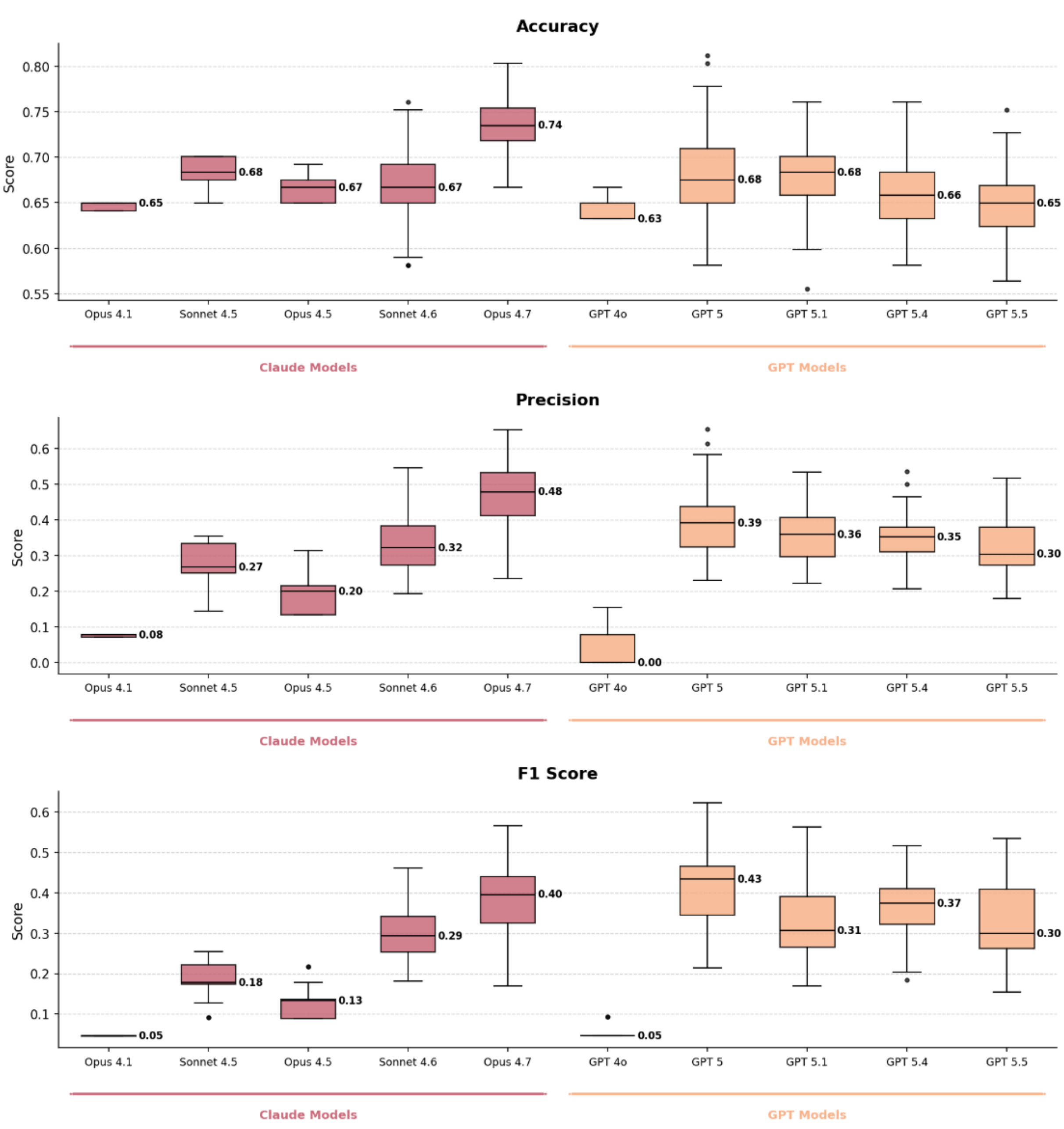

**Figure 7.** Aggregate Performance Rates Across Generations and Models

Precision and F1 follow a similar pattern but with notable inconsistencies: Claude Opus 4.1 produces near-zero median Precision (0.08) and F1 (0.05), while Claude Opus 4.5 improves to a median Precision of 0.20 and F1 of 0.13. Claude Sonnet 4.5 and Claude Sonnet 4.6 continue the improvement to median Precision values of 0.27 and 0.32 and median F1 values of 0.18 and 0.29, respectively. Claude Opus 4.7 achieves the strongest aggregate performance within the Claude family, with a median Precision of 0.48 and a median F1 of 0.40. Notably, however, Claude Opus 4.5 underperforms Claude Sonnet 4.5 in both Precision and F1 despite being a later Opus variant.

For GPT models, the generational trend is considerably less consistent. GPT-4o exhibits near-zero median Precision (0.00) and a median F1 of just 0.05, reflecting severely limited detection capability. GPT-5 represents a substantial leap, achieving a median Accuracy of 0.68, a median Precision of 0.39, and the highest median F1 across all GPT variants at 0.43. However, subsequent versions fail to sustain this improvement: GPT-5.1 produces a median F1 of 0.31, GPT-5.4 of 0.37, and GPT-5.5 of 0.30, all below GPT-5, despite representing later iterations. Accuracy similarly fluctuates, with GPT-5.4 (0.66) and GPT-5.5 (0.65) falling below GPT-5.1 (0.68) and GPT-5 (0.68).

The spread and asymmetry of the aggregate metric distributions further emphasize execution-level instability within each family. Within the Claude family, Claude Opus 4.7 exhibits the widest Precision and F1 spreads (ranges of 0.42 and 0.40, std of 0.089 and 0.080), reflecting substantial execution-level variation even at peak performance. Earlier Claude models, such as Opus 4.1, show near-zero spread (F1 std=0.001), indicating collapsed rather than stable behavior. Among GPT models, GPT-5 and GPT-5.1 exhibit the widest F1 distributions (ranges of 0.41 and 0.39, respectively, with a standard deviation of 0.085 and 0.092, respectively), with extended upper whiskers indicating that strong balanced performance is achievable but far from consistent across executions. GPT-5.5 similarly shows high F1 variability (std=0.096) despite a lower median.

Thus, although later models achieve stronger performance, the improvement trend is not monotonic, and they do not necessarily provide more stable requirements evaluation capability.

### 4.4 Sensitivity Analysis

To examine how far the findings generalize beyond the primary experimental setup, we conducted two sensitivity probes. The first compares performance across two requirement sets, including a lower-quality set with more explicit issues. The second examines sampling temperature, which controls the stochasticity of LLM outputs. The findings from the sensitivity probes are discussed below.

#### *4.4.1 Sensitivity Probe 1: Requirement Set Robustness*

To examine whether the non-monotonic generational trends observed previously are specific to a single requirement set, the same cross-generational comparison was conducted using Requirement Set B. Figures 8 and 9 present the comparison of normalized and aggregated metrics between GPT and Claude variants, respectively. In these figures, in addition to the box plots showing the distribution of metrics across repeated runs for each model under Set B, dashed blue lines mark the corresponding Set A median for each model, marking the baseline values from Section 4.3. Blue labels are shown only when the Set A and Set B medians differ by more than 0.025 to avoid cluttering. Hence, an unlabeled blue dashed line indicates the two medians are almost equivalent.

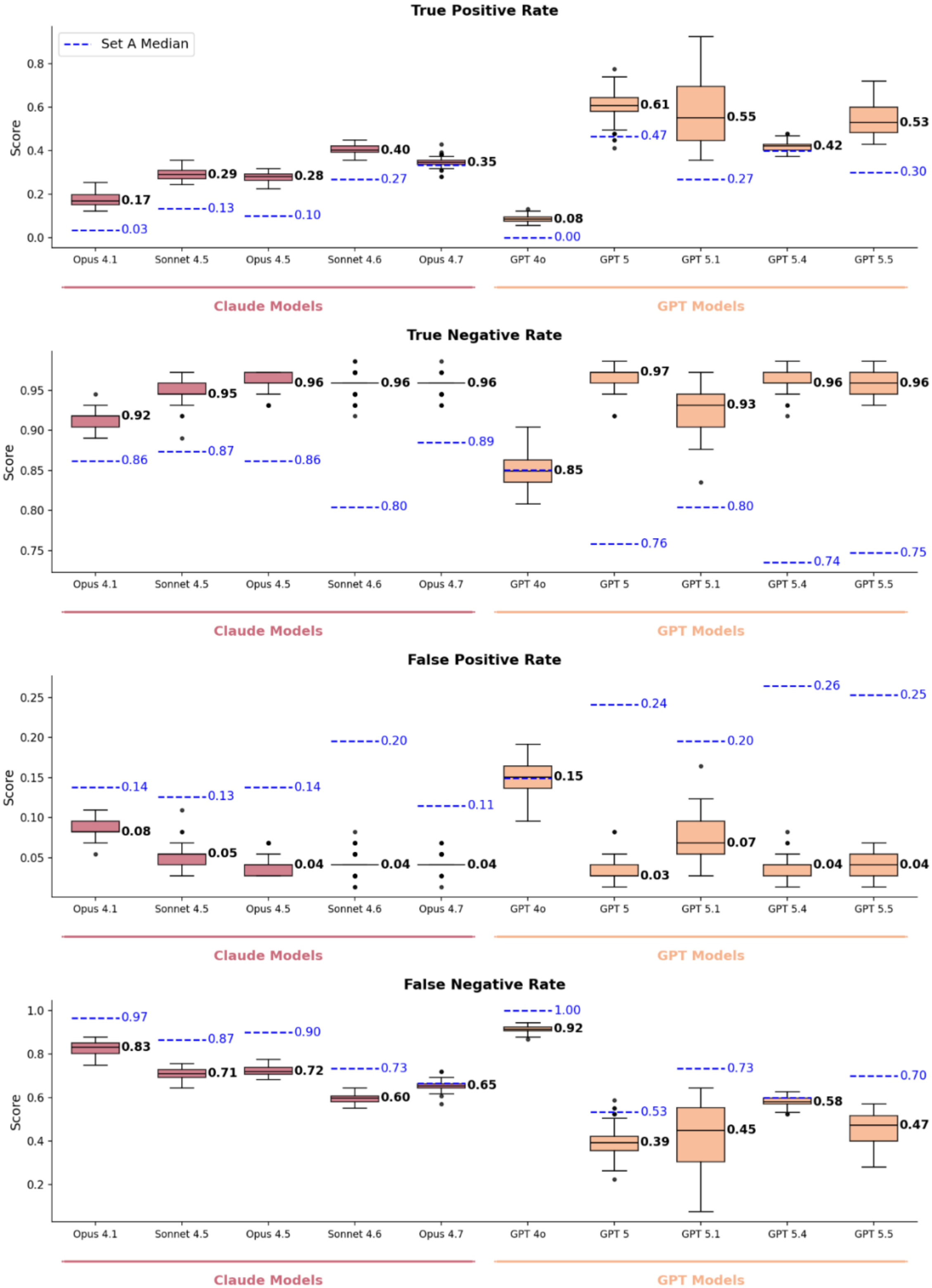

**Figure 8.** Comparison of Normalized Performance Rates Across Generations and Models for Requirements Set B

The figures show that the overall generational performance patterns remain highly consistent with the earlier findings. Within the Claude family, TPR improves from 0.17 for Claude Opus 4.1 to 0.40 for Claude Sonnet 4.6, but subsequently decreases to 0.35 for Claude Opus 4.7 despite the latter being a newer variant. Similarly, within the GPT family, GPT-5 achieves the highest median TPR at 0.61 and the strongest F1-score at 0.75, while later variants such as GPT-5.4 and GPT-5.5 decline to TPR values of 0.42 and 0.53 and F1 values of 0.58 and 0.68, respectively. GPT-5.1 also demonstrates weaker and substantially more variable detection performance than GPT-5 despite being a subsequent generation. These fluctuations indicate that improvements across generations are again inconsistent and do not uniformly favor later model releases.

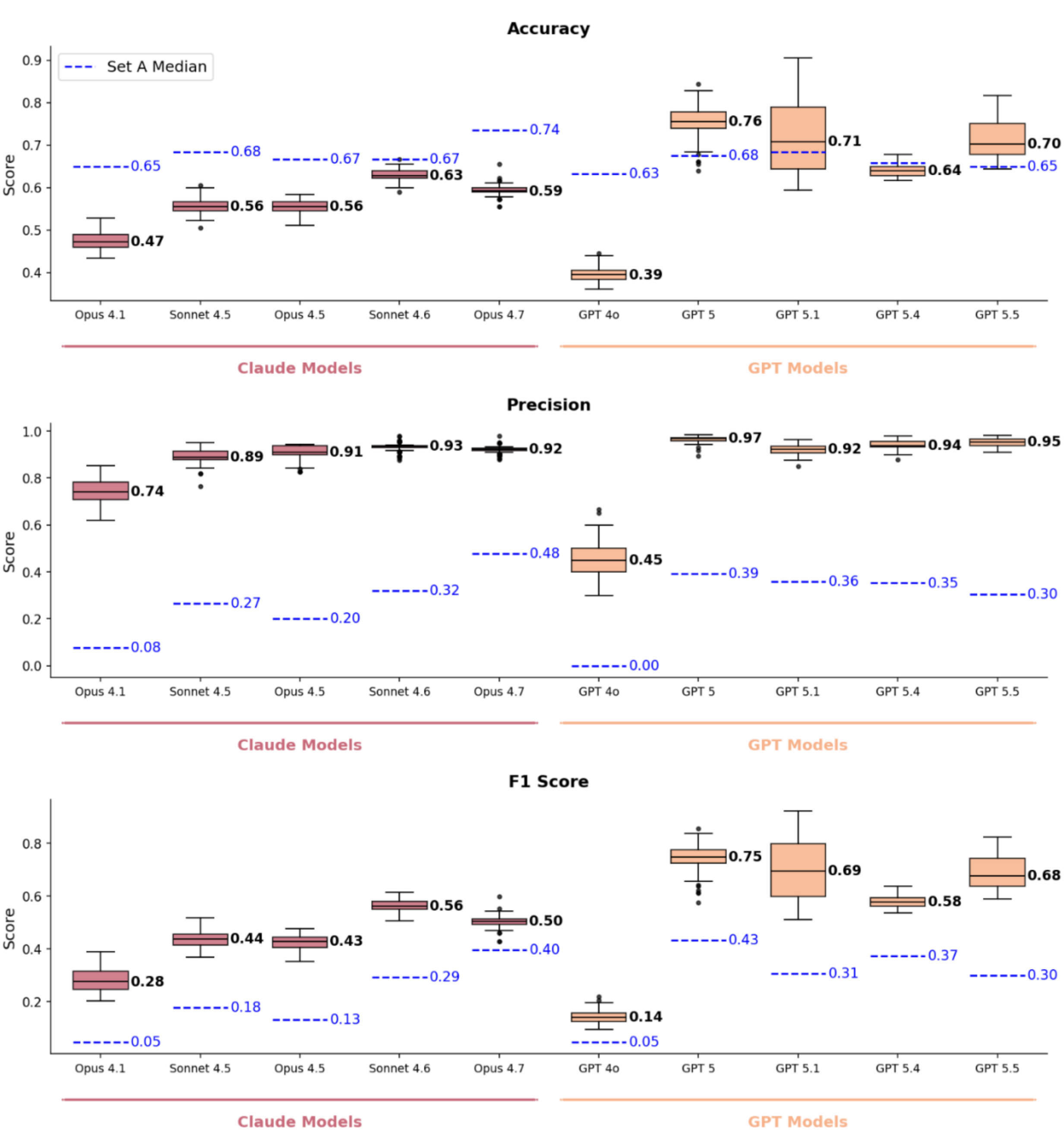

**Figure 9.** Comparison of Aggregate Performance Rates Across Generations and Models for Requirements Set B

The distributional characteristics further indicate that execution-level instability and fluctuating performance are not unique to the original requirement set. For example, within the Claude family, variability in TPR distributions remains comparatively contained yet fluctuating: Claude Opus 4.1 exhibits a TPR range of 0.12–0.25 (std=0.029), widening to 0.22–0.32 (std=0.021) for Claude Opus 4.5, 0.24–0.36 (std=0.027) for Claude Sonnet 4.5, 0.36–0.45 (std=0.020) for Claude Sonnet 4.6, and 0.28–0.43 (std=0.021) for Claude Opus 4.7. Notably, Claude Opus 4.1 has the highest standard deviation, suggesting that distributional instability is not strictly tied to generation. Within the GPT family, the picture is considerably more volatile. GPT-5.1 stands out as the most unstable model across both datasets, exhibiting a TPR range of 0.36–0.93 with a standard deviation of 0.153 and an F1 range of 0.51–0.92 (std=0.113). GPT-5.5 also exhibits substantial TPR variability (0.43–0.72, std=0.070) and F1 variability (0.59–0.82, std=0.058), while GPT-5.4 remains comparatively stable (TPR std=0.021, F1 std=0.021), which is in alignment with the non-monotonic variability pattern observed in the original dataset.

Another important pattern emerges from this sensitivity analysis. The median model performance under Set B, which is a poorer set compared to Set A, exceeds performance observed under Set A for most evaluation metrics. For example, across all ten models, median TPR is higher under Set B than under Set A, this effect is most pronounced for GPT-5, which increases from 0.47 on Set A to 0.61 on Set B, and Sonnet 4.6, which increases from 0.27 to 0.40. A similar pattern is observed for TNR, where median values under Set B are higher than or approximately equal to those under Set A for every model. The largest margins are observed for GPT-5, which increases from 0.76 to 0.97, and GPT-5.4, which increases from 0.74 to 0.96. Complementary error-rate metrics support this interpretation, with lower missed-detection and false-alarm tendencies under Set B.

The largest differences between the two requirement sets occur for precision and F1 score. Every model performs better under Set B on both metrics, indicating that LLM predictions are not only more likely to detect issues, but also more reliable when issues are flagged in this lower-quality requirement set. Precision gains are particularly large for Sonnet 4.6, which increases from 0.32 on Set A to 0.93 on Set B, and GPT-5, which increases from 0.39 to 0.97. F1-score gains are similarly pronounced for GPT-5.1, which increases from 0.31 to 0.69, and Sonnet 4.6, which increases from 0.29 to 0.56. Accuracy is the only exception to this trend. All five Claude models and GPT-4o exhibit lower accuracy under Set B than under Set A. For example, Opus 4.1 decreases from 0.65 to 0.47, Sonnet 4.5 from 0.68 to 0.56, and GPT-4o from 0.63 to 0.39. By contrast, the later GPT models maintain or improve accuracy under Set B; for example, GPT-5 increases from 0.68 to 0.76, and GPT-5.1 increases from 0.68 to 0.71.

Overall, this sensitivity analysis supports the earlier findings in Section 4.3 that suggests that the newer LLM generations do not inherently guarantee superior and more reliable requirement quality evaluation performance. Moreover, these results suggest that off-the-shelf LLMs may perform better when requirement quality issues are highly visible, but their performance declines in the more representative SE context.

#### *4.4.2 Sensitivity Probe 2: Temperature Robustness*

Figure 10 presents the sensitivity of the requirement quality evaluation performance to sampling temperature using GPT 5.4 on Requirement Set A. The results show that both median performance and execution-level variability remain relatively stable across the examined temperature settings, with only modest fluctuations and no consistent or monotonic trend. We discuss Figure 10 in detail below.

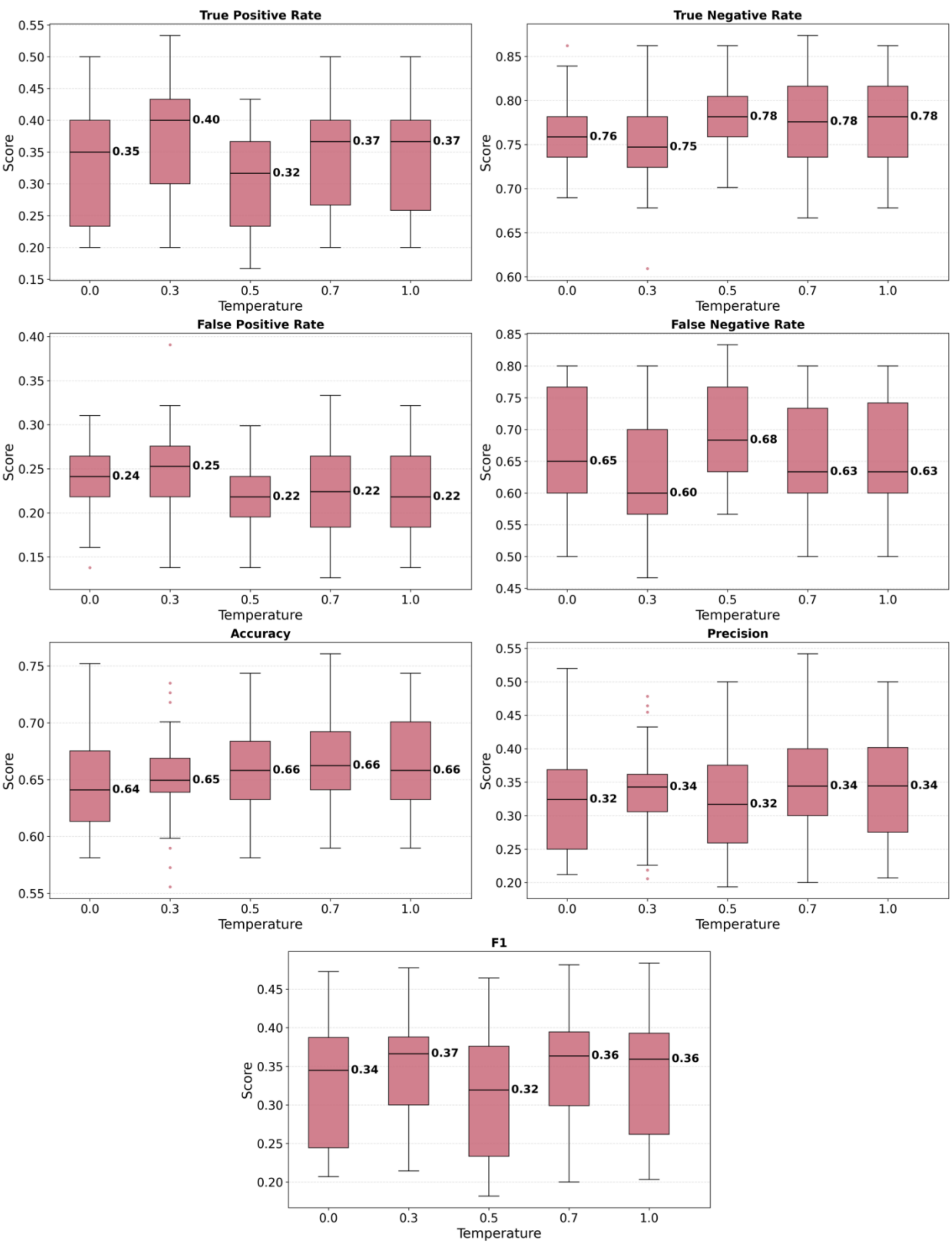

**Figure 10.** Distribution of Performance Metrics Across Sampling Temperatures

The median TPR increases from 0.35 at temperature 0.0 to 0.40 at 0.3, declines to 0.32 at 0.5, and then increases to 0.37 at both 0.7 and 1.0. The complementary median FNR exhibits the inverse pattern, changing from 0.65 at temperature 0.0 to 0.60 at 0.3, increasing to 0.68 at 0.5, and then decreasing to 0.63 at 0.7 and 1.0. TPR standard deviations are between 0.08 and 0.09, with corresponding ranges of 0.20–0.50, 0.20–0.53, 0.17–0.43, 0.20–0.50, and 0.20–0.50 at temperatures 0.0, 0.3, 0.5, 0.7, and 1.0, respectively. FNR exhibits the corresponding variability, with ranges of 0.50–0.80, 0.47–0.80, 0.57–0.83, 0.50–0.80, and 0.50–0.80 across the five temperature settings. Thus, neither increasing nor decreasing temperature monotonically improves the model's ability to detect existing requirement quality issues or reduces variability in its detection performance. A similar non-monotonic pattern is observed in the model's ability to avoid false alarms. Median TNR varies from 0.76 at temperature 0.0 to 0.75 at 0.3, increases to 0.78 at 0.5, 0.7, and 1.0. Variability also changes irregularly. TNR standard deviations are 0.04, 0.04, 0.04, 0.05, and 0.04 at temperatures 0.0 through 1.0, respectively, while the corresponding ranges are 0.69–0.86, 0.61–0.86, 0.70–0.86, 0.67–0.87, and 0.68–0.86. The complementary median FPR follows the inverse pattern. Thus, temperature change is not consistently associated with either better false-alarm avoidance or greater execution-level variability.

Aggregate performance metrics reinforce this pattern. Median Accuracy varies from 0.64 at temperature 0.0 to 0.65 at 0.3, 0.66 at 0.5, 0.66 at 0.7, and 0.66 at 1.0. Its ranges are 0.58–0.75, 0.56–0.74, 0.58–0.74, 0.59–0.76, and 0.59–0.74, respectively, with standard deviations between 0.03 and 0.04. Precision similarly changes non-monotonically, with median values of 0.32, 0.34, 0.32, 0.34, and 0.34 across temperatures 0.0, 0.3, 0.5, 0.7, and 1.0. Precision standard deviations are 0.07, 0.05, 0.08, 0.07, and 0.07 with corresponding ranges of 0.21–0.52, 0.21–0.48, 0.19–0.50, 0.20–0.54, and 0.21–0.50 at temperatures 0.0 through 1.0, respectively. F1-score shows more noticeable differences in median performance, increasing from 0.34 at temperature 0.0 to 0.37 at 0.3, declining to 0.32 at 0.5, recovering to 0.36 at 0.7, and remaining similar at 0.36 at 1.0. Its execution-level variability changes slightly without a consistent temperature-dependent pattern. F1-score standard deviations are 0.08, 0.06, 0.07, 0.07, and 0.07 across increasing temperatures, with corresponding ranges of 0.21–0.47, 0.21–0.48, 0.18–0.46, 0.20–0.48, and 0.20–0.48.

Overall, these results indicate that sampling temperature has limited influence on requirement quality evaluation performance. Moreover, execution-level variability persists even in the lower temperatures, suggesting that temperature tuning alone cannot ensure deterministic evaluation behavior.

## 5. Discussion

Advances in generative AI have stimulated interest in using LLMs to support RE activities. However, requirement quality evaluation is not simply a language-processing task; it requires contextual reasoning, systems-level judgment, and disciplined application of engineering quality criteria. To that end, the objective of this study was to characterize the capability of off-the-shelf LLMs to evaluate the quality of requirements for engineered systems. We investigated overall evaluation performance, behavior across INCOSE quality criteria, cross-generational performance trends, and the influence of sampling temperature. The results reveal several patterns: off-the-shelf LLMs are better at avoiding false alarms than detecting existing quality issues, performance varies substantially across quality criteria, newer model generations do not consistently outperform earlier ones, and temperature variation has limited influence on evaluation performance. The following discussion synthesizes these findings. Together, these findings suggest that for appropriate integration into SE workflows, off-the-shelf LLMs should be viewed not as autonomous requirement quality evaluators in their current capability but as decision-support tools whose value depends on careful human oversight.

### 5.1 Missed Detections Are the Dominant Risk in LLM-Based Requirement Evaluation

The performance results reveal that off-the-shelf LLMs are more likely to miss real requirement quality issues than to overburden reviewers with false alarms. Across the evaluated off-the-shelf LLM models, FNRs were consistently substantially higher than FPRs indicating that the dominant error mode is missed detection. From an SE perspective, this is the more consequential failure mode because missed defects can propagate into design, verification, and validation before they are discovered. Such missed defects may ultimately result in under- or over-designed systems, both of which contribute to increased engineering effort, schedule delays, and lifecycle cost. Although the FPR is comparatively lower, it is not negligible. Incorrectly identifying valid requirements as problematic may prompt unnecessary requirement revisions, design modifications, and verification activities, thereby increasing SE effort without improving system quality.

Consequently, the low median F1-score indicates that the model fails to achieve a reliable balance between issue detection and prediction reliability. Such behavior limits the utility of off-the-shelf LLMs as autonomous evaluators and particularly concerning from a trust perspective. An evaluator that misses two-thirds of real issues while simultaneously producing incorrect flags in nearly half of its positive predictions offers little basis for confident reliance. Over time, exposure to such errors is likely to undermine practitioner trust and limit the adoption of AI-assisted requirement evaluation in practice (Kahr et al., 2024; Yang et al., 2023). In addition, none of the metrics exhibits perfect performance in any of the executions, indicating that evaluation errors persist even in the most favorable cases. These findings suggest that current off-the-shelf LLMs cannot yet be considered dependable autonomous requirement quality evaluators. However, they may still be useful in a human-in-the-loop workflow, particularly by reducing reviewer effort by surfacing a subset of problematic requirements early.

Another interesting takeaway is that off-the-shelf LLMs performed better when requirement quality issues are more salient, explicit, or frequent, but their performance declines in the more representative SE requirements Set A. This is concerning for SE practice, because real-world requirement evaluation often requires identifying subtle, context-dependent issues rather than only detecting highly visible issues. Therefore, the weaker performance observed on Set A may better reflect the level of performance SE practitioners should expect when applying off-the-shelf LLMs to more realistic requirement evaluation contexts.

Beyond performance trends, the findings show notable variability in LLM behavior across executions, even when the underlying model, prompt structure, and evaluation criteria remain unchanged. For practitioners, this variability introduces uncertainty regarding the trustworthiness of any individual LLM-generated assessment, as there is no clear indication whether a particular execution reflects stronger or weaker model performance.

### 5.2 Requirement Quality Criteria Are Not Equally Suitable for LLM-Based Evaluation

This criterion-level analyses have an important implication for AI-assisted RE: requirement quality evaluation performance is not uniform across requirement quality criteria. This is unsurprising given that different criteria require different forms of reasoning in SE practice. Criteria such as Unambiguous and Conforming exhibit comparatively better performance in terms of avoiding missed issue detection. Arguably, issues in these criteria are more directly observable through local textual analysis, such as vague terminology, missing qualifiers, violating organizational requirement-writing standards and formatting norms. In contrast, quality issues related to categories such as Correct, Necessary, and Complete often go undetected by the LLM. Assessing Correctness and Necessity requires judgment that depends on integrating information across the problem statement, ConOps, and broader SE knowledge. Poor performance in these categories suggests that the LLM struggles substantially to establish these contextual and purpose-driven linkages, which are essential for successful SE. This limitation indicates LLM's lack of domain-specific engineering judgment and the absence of broader situational understanding that human experts routinely apply when evaluating the requirements. In practice, missed issues in these criteria can have severe downstream consequences. Requirements that are incorrect, unnecessary, or incomplete may guide design teams toward omitted or unnecessary system capabilities. These effects can exacerbate existing shortcomings in SE practice by increasing rework, verification burden, schedule delays, and lifecycle cost.

Turning to false alarm behavior, the criterion-level analyses support the earlier discussion on overall better performance in avoiding false alarms compared to issue detection. Particularly, Appropriate, Correct, and Unsure of Category demonstrate near-zero FPRs, meaning that false alarms almost never occur in these categories. It is worth noting that the LLM never assigned "Unsure of Category" to any requirement across all evaluation attempts, which is aligned with the ground truth. This indicates that the LLM reliably avoids incorrectly introducing uncertainty classifications in requirement evaluation when no uncertainty exists in the input data. No Issues, Complete, Necessary, and Singular similarly exhibit relatively strong tendency to avoid false alarms. Interestingly, this behavior is contradictory to the FNR results for Correct and Necessary. This suggests that primary failure mode for these categories is missed detection rather than false alarm generation. In contrast, categories such as Unambiguous and Conforming exhibit elevated false alarms. This suggests that stronger detection in these categories is partially achieved at the cost of increased false alarms. From an SE perspective, these false alarms can still be costly because they may result in unnecessary requirement revisions, design changes, verification effort, and review cycles without improving system quality.

More broadly, different reasoning demands in different quality criteria give rise to different failure modes in LLM-based requirement quality evaluation. This contrast in different failure modes across different metrics suggests that stronger issue detection should not be interpreted as better quality judgment unless it is accompanied by reliable false-alarm control. Importantly, not all requirement quality criteria are equally well suited to LLM-based evaluation. For example, criteria such as ambiguity and conformance may be better supported through other approaches, such as rule-based, ontology-backed approaches (Chen et al., 2020; Dermeval et al., 2016).This suggests that different requirement quality criteria may require targeted mitigation strategies that address the specific reasoning limitations underlying each failure mode.

### 5.3 Newer LLM Generations Do Not Necessarily Yield Better Requirement Evaluation Performance

The cross-generational comparison of LLM performance further reveals that newer LLM generations do not monotonically outperform earlier ones in requirement quality evaluation despite rapid model evolution and increasing model complexity. In several cases, newer models exhibit performance tradeoffs, achieving improvements in some metrics while demonstrating weaker performance in others, thus failing to consistently outperform their predecessors. For example, compared with Opus 4.5, Sonnet 4.6 improved TPR but showed higher FPR, indicating stronger issue detection at the cost of more false alarms. Similarly, GPT-5.5 showed higher TNR relative to GPT-5.4, but this improvement was accompanied by decreases in accuracy and precision. Furthermore, the findings reveal that later generations also exhibit wider distributional spreads and greater execution-to-execution variability in many instances. These findings indicate that generational advancement alone is not a reliable predictor of requirement quality evaluation capability. One possible explanation is that general-purpose LLM improvements are not specifically optimized for requirement quality assessment.

For SE organizations, this means that model adoption decisions should not be based on general frontier-model rankings or assumptions that the newest model will necessarily provide better requirement quality judgments. Instead, greater value may be obtained by focusing on task-specific adaptation strategies, such as prompt engineering, contextual augmentation, domain-specific knowledge integration, or model customization, to better align LLM behavior with the RE task at hand.

### 5.4 Temperature Tuning Does Not Guarantee Improved Evaluation Performance or Stability

The temperature sensitivity analysis produced a notable finding. The results indicate that requirement quality evaluation performance remains stable within the examined sampling temperature range. Although both median performance and execution-level variability fluctuate across temperature settings, the magnitude of these differences is small and no consistent or monotonic relationship with temperature is observed. Thus, within the examined range, adjusting sampling temperature neither substantially changes the model's typical evaluation performance nor provides a reliable means of controlling execution-level variability. Furthermore, the persistence of substantial variability at lower temperature settings, such as, 0.0 and 0.3, across all performance metrics demonstrates that reducing sampling temperature does not guarantee deterministic evaluation behavior. This result is particularly surprising given the widespread perception that temperature serves as a mechanism control LLM stochastic behavior. Higher temperatures often associated with greater creativity, diversity, and variability, and lower temperatures with more deterministic behavior (Agarwal et al., 2024; Bellemare-Pepin et al., 2026; Wang et al., 2026).

From a practical perspective, these findings indicate that temperature tuning alone is unlikely to provide a reliable means of improving requirement quality evaluation performance or ensuring stable evaluation outcomes. Rather, performance appears to be governed primarily by the model's underlying reasoning capabilities, SE and RE knowledge, and ability to establish the contextual relationships necessary for quality assessment. Consequently, practitioners are likely to obtain greater performance gains through targeted interventions in improving these capabilities than through temperature tuning.

### 5.5 Limitations

This study has several limitations that should be considered when interpreting the findings. First, the analysis is based on a smaller scale system (i.e., a point mass manipulator) relative to larger engineered systems such as aircraft, satellites, automobiles, or smart devices. Although this base case still needs a strong coupling between hardware and software elements to fulfill system level objectives; this limited scope may limit the generalizability of the results to other domains, requirement maturity levels, and systems with greater technical or organizational complexity. Nevertheless, as greater system complexity would likely increase the contextual and systems-level reasoning demands placed on LLMs, these findings may be interpreted as a conservative benchmark of off-the-shelf LLM performance for requirement quality evaluation. The documented issues are only expected to worsen with higher scope and complexity.

Second, the study evaluates off-the-shelf LLMs under a defined prompting and evaluation setup. Prior work suggests that advanced prompting strategies and specialization techniques, such as few-shot prompting, retrieval augmentation, fine-tuning, domain-specific instruction tuning, structured reasoning prompts may improve performance (Chung et al., 2024; Dodgson et al., 2023; Fazelnia et al., 2024; Le et al., 2025; Sivarajkumar et al., 2024). As such, our findings should be interpreted as a documentation of the lower bound of LLM performance as the purpose of this study was not to optimize model performance but to but to demonstrate where current off-the-shelf LLMs stand so that future capability development can be better targeted and assessed.

Third, the study does not exhaustively evaluate all versions of all available LLMs, and the temperature sensitivity analysis considers a limited range of temperature values. This scope choice also excludes locally hosted, closed-loop, or restricted models, whose performance may differ depending on training data, deployment context, and access to organization-specific knowledge. Finally, requirement quality assessment is operationalized as a binary classification task, which does not capture severity, rationale quality, or partial correctness of LLM judgments. Hence, while the findings documented in this study provide valuable insight into the behavior of off-the-shelf LLMs, the specific performance values should not be interpreted as universally representative of all requirement quality assessment settings and contexts. Nonetheless, although the magnitude of performance, error rates, and variability may differ across systems, models, and prompting strategies, the broader findings are expected to remain relevant for AI-assisted requirement assessment in engineered systems.

## 6. Conclusion

This study characterized the capability of off-the-shelf LLMs to evaluate the quality of requirements for engineered systems using expert-derived ground truth and INCOSE quality criteria. The insights from this study have several important implications for AI4RE practitioners and researchers. The findings indicate that the current capabilities of off-the-shelf LLMs are far from supporting fully autonomous requirement quality evaluation in a trustworthy manner as their performance remains constrained by substantial missed detection rates, false alarms, criterion-specific failure modes, and execution-level variability. These findings suggest the need for a human-in-the-loop review process consistent with prior work advocating human-AI collaboration in design and decision-making (Gyory et al., 2022; Song et al., 2022b; Viros i Martin & Selva, 2022). Rather than replacing human expert judgment, LLM-based evaluation tools should be viewed as decision-support mechanisms that reinforce human expertise while preserving human authority over final quality assessments. Furthermore, the substantial execution-level variability observed across models suggests that single-execution evaluations are statistically unreliable. This raises important questions regarding how LLM-based requirement evaluations should be aggregated, validated, and presented to practitioners, particularly for high-stakes requirement sets.

Furthermore, findings suggest that newer model generations do not consistently outperform their predecessors, and temperature tuning does not provide a reliable mechanism to improve evaluation performance or reduce variability. Therefore, future AI4RE research should explore approaches that address more fundamental challenges related to reasoning, contextual understanding, and engineering judgment with targeted interventions, including prompt engineering, few-shot examples grounded in INCOSE quality criteria, or retrieval-augmented context provision. The criterion-level results further point to Correct and Necessary as priority targets for such interventions, given their near-zero detection. Additionally, given the observed variability in LLM evaluation performance, further research is needed to develop human–AI collaborative workflows that effectively balance automated support with expert judgment. Collectively, these efforts support a vision of AI-assisted requirements evaluation that augments, rather than replaces, systems engineers while enhancing transparency, robustness, and trustworthiness.

**Acknowledgments:** This research was supported by the National Nuclear Security Administration (NNSA) through Systems Engineering Research Center (SERC) Contract WRT-2416: Systems Engineering Beyond the Horizon. Any views, opinions, findings and conclusions or recommendations expressed in this material are those of the authors and do not necessarily reflect the views of NNSA.

**Data Availability Statement:** The data that support the findings of this study are available from the corresponding author upon reasonable request.

**Conflict of interest:** The authors declare no conflicts of interest.

**Appendix A**

Appendix A contains the system prompt used in this research.

```
You are a requirements quality evaluator. You will receive a list of requirements
with their IDs, along with Project Context, Concept of Operations (ConOps)
information, and a reference image of the ConOps.

When completing the following tasks, carefully consider the provided Project
Context, Concept of Operations, and reference image to determine if requirements
align with the project goals, needs, and operational constraints.

Your task is to perform evaluations as instructed below.

## Task: INDIVIDUAL REQUIREMENT EVALUATION

Evaluate each requirement individually against ALL of these quality criteria (A1-
A9). Be CRITICAL and THOROUGH in identifying issues.

IMPORTANT: Criterion A1 means NO violation of any of the criteria for a requirement.
If there are no violations of any criteria for the requirement, mark it.

Criteria A2-A9 mean violation of that criterion EXISTS. For each requirement,
include all the criteria that are violated.

Be thorough and critical - if there is an issue, mark it. If issues exist, return
one object per issue criterion, including:
•     criterion ID
•     criterion name
•     short explanation on why the issue exists (1–2 sentences)

The criteria are as follows:
A1- No issues: No issues found with the requirement (mark ONLY if the requirement
has NO problems)
A2-Necessary: The requirement statement defines a capability, characteristic,
constraint, or quality factor needed to satisfy a life cycle concept, need, source,
or parent requirement.
A3-Appropriate: The specific intent and amount of detail of the requirement
statement is appropriate to the level (e.g., the level of abstraction, organization,
or system architecture) of the entity to which it refers.
A4-Unambiguous: The requirement statement is stated such that the intent is clear
and the requirement can be interpreted in only one way by all the intended
stakeholders.
A5-Complete: The requirement statement sufficiently describes the necessary
capability, characteristic, constraint, conditions, or quality factor to meet the
need, source, or higher-level requirement from which it was transformed.
A6-Singular: The requirement statement states a single capability, characteristic,
constraint, or quality factor.
A7-Correct: The requirement statement is an accurate representation of the need,
source, or higher-level requirement from which it was transformed.
A8-Conforming: The requirement statement conforms to an approved standard pattern
and style guide or standard for writing and managing requirements.
A9-Unsure of Category (ISSUE EXISTS if you cannot categorize the problem)
```

```
## OUTPUT FORMAT

Return ONLY valid JSON.

{
  "individualEvaluations": {
    "FR.1": [
      {
        "criterion": "A4",
        "name": "Unambiguous",
        "explanation": "The requirement uses subjective wording that allows multiple interpretations."
      }
    ],
    "FR.2": []
  }
}
```

**Appendix B**

As supplementary analysis to criterion-level error rates discussed in Section 4.2, Appendix B presents the distributions of TPR, TNR, accuracy, precision, and F1 at criterion level using Claude Opus 4.7. The brief description of criterion-level analysis of each of these metrics are provided below.

**Appendix B.1. True Positive Rate (TPR) Per Quality Criterion**

Figure B1 presents the distribution of TPR across individual requirement quality issue categories. Overall, the results show that issue detection varies substantially by category. The weakest detection performance occurs for Correct and Necessary. For Correct, TPR remains close to zero across executions, with a median of 0.11 and a mean of 0.09, showing that correctness-related issues are almost always missed. Detection of Necessary represents the most severe failure case: TPR remains 0.00 across all executions, indicating that the model fails to detect any necessity-related issue throughout the evaluation.

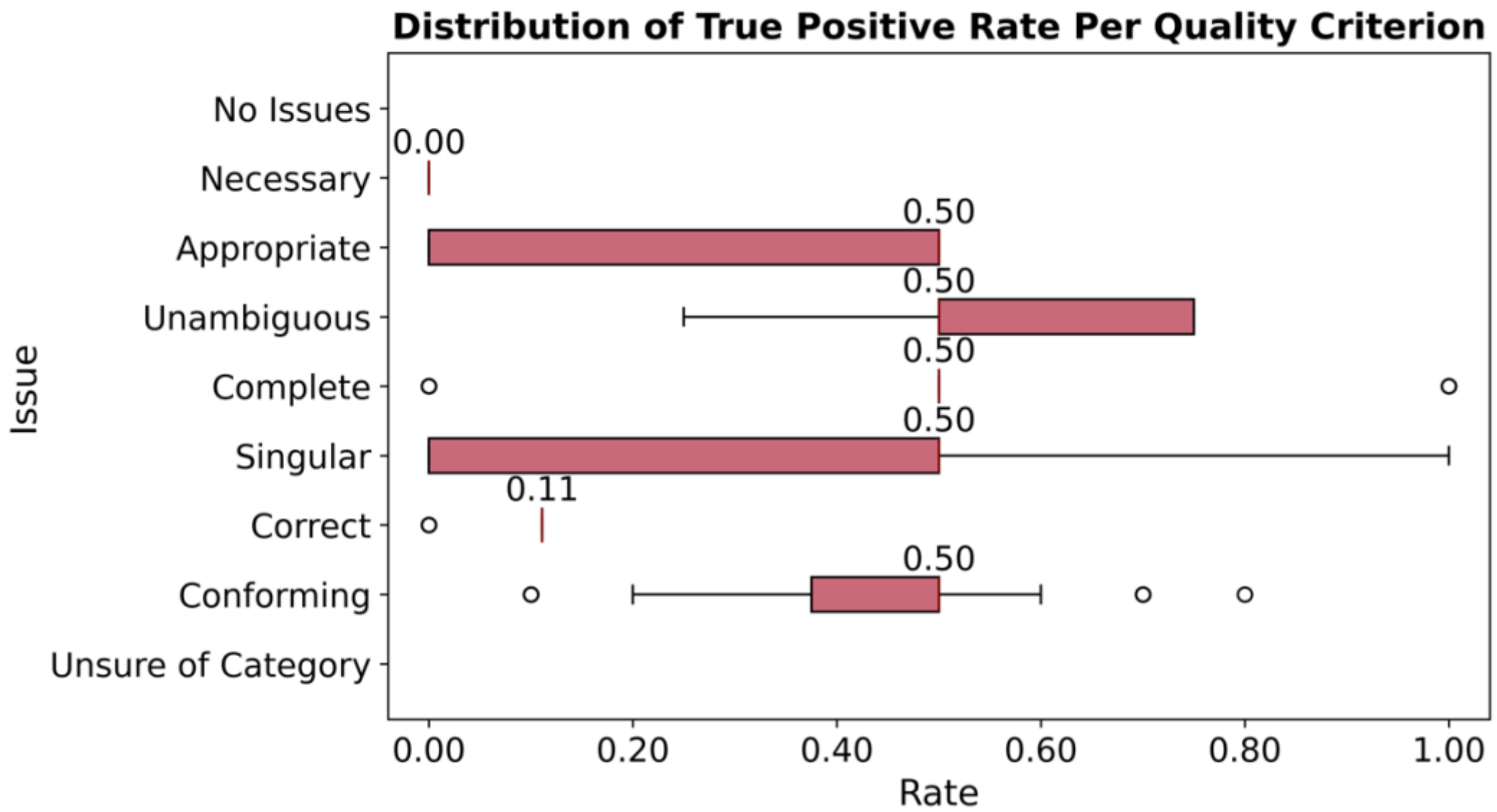

**Figure B1.** TPR Distribution by Quality Criterion

Detection performance improves but remains concerning for Appropriate, Singular, and Complete. Although each category has a median TPR of 0.50, their distributions show substantial instability. Detection of Appropriate has a lower mean TPR of 0.29, suggesting that many executions detect few appropriateness-related issues. Detection of Singular spans the full 0.00 to 1.00 range, indicating highly unpredictable detection behavior. Detection of Complete shows the most unstable behavior among these mid-performing categories, with TPR ranging from 0.00 to 1.00 and the largest standard deviation, indicating no dependable detection pattern.

Among the less severe failure cases, detection of Unambiguous and Conforming show moderate detection performance, each with a median TPR of 0.50. Detection of Unambiguous is comparatively more stable, with TPR ranging from 0.25 to 0.75 and a mean of 0.57. Conforming also centers around a median TPR of 0.50, but shows a wider range from 0.10 to 0.80, indicating more variation across executions. Finally, No Issues and Unsure of Category are omitted from the TPR analysis because no true positive instances exist in the ground truth for these categories, making TPR mathematically undefined.

**Appendix B.2. True Negative Rate (TNR) Per Quality Criterion**

Figure B2 presents the distribution of TNR across issue categories, showing how reliably the LLM correctly identifies non-issues. Overall, TNR values are substantially higher than the TPR values reported in Figure B1, indicating that the model is generally stronger at avoiding false alarms than detecting actual quality issues.

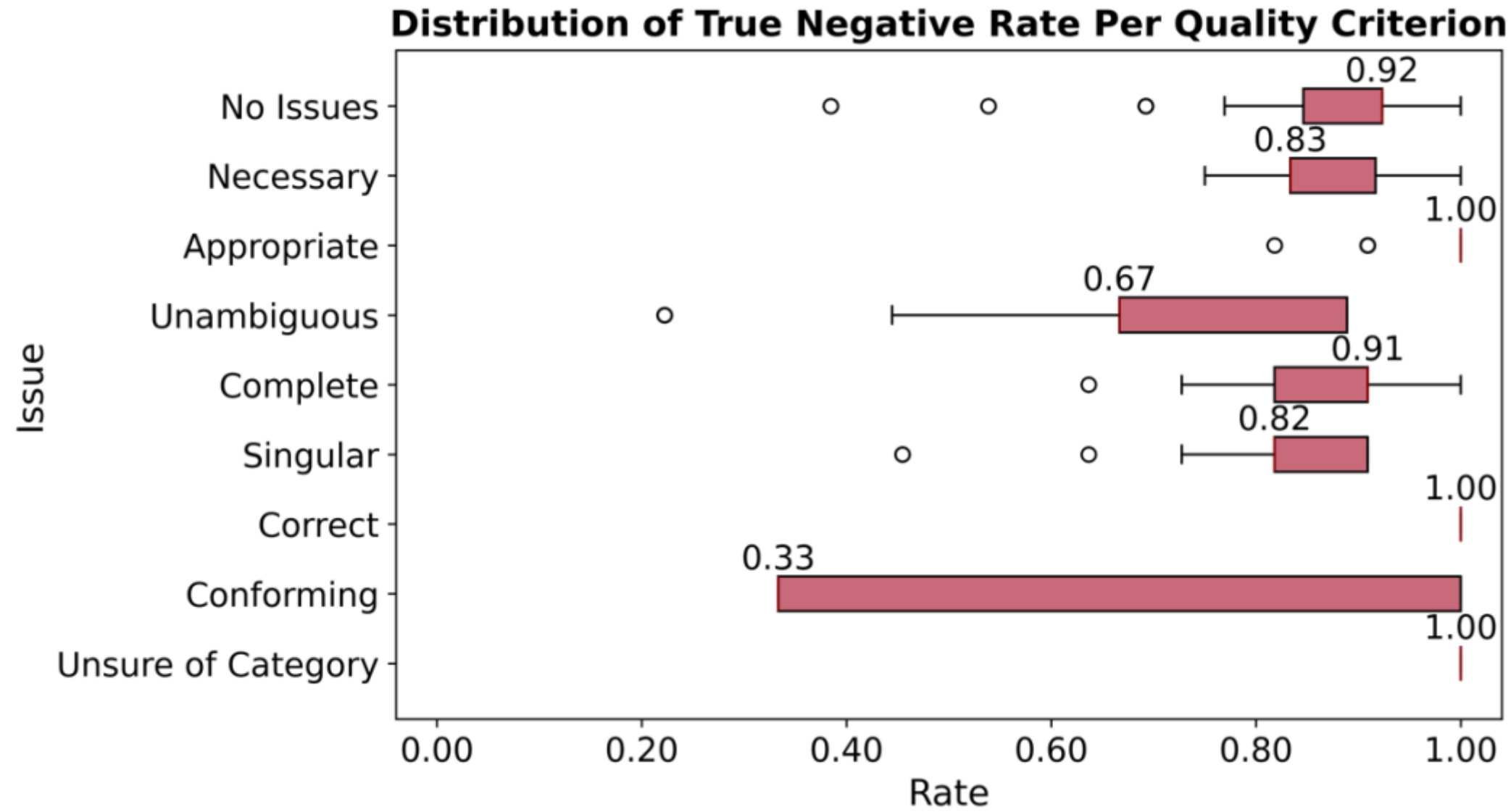

**Figure B2.** TNR Distribution by Quality Criterion

The weakest TNR performance occurs for Conforming, where non-issue classification is highly inconsistent. TNR ranges from approximately 0.33 to 1.00, with a median of 0.33 and a mean of 0.55, indicating that the model frequently misclassifies valid conforming requirements as problematic. Unambiguous also shows weaker non-issue classification, with TNR ranging from approximately 0.22 to 0.89, a median of 0.67, and a mean near 0.70, suggesting elevated variability and frequent false alarms.

Performance improves for Singular, Necessary, and Complete. Singular shows a median TNR of 0.82, Necessary 0.83, and Complete 0.91, with relatively low variability across executions. These results suggest that the model is generally more stable when rejecting non-issues in these categories. The strongest performance is observed for No Issues, Appropriate, Correct, and Unsure of Category. No Issues shows a median TNR of 0.92, while Appropriate, Correct, and Unsure of Category exhibit collapsed TNR distributions at 1.00, indicating that the model consistently avoids false alarms for these categories.

**Appendix B.3. Accuracy Per Quality Criterion**

Figure B3 presents the distribution of Accuracy across requirement quality criteria, providing an aggregate view of the LLM’s overall performance by accounting for both the successful detection of issues and the correct rejection of non-issues. The weakest performance is observed for Correct and Conforming. Accuracy for Correct remains consistently low, with a median of 0.38, mean of 0.37, and very small variability, indicating stable but poor classification performance across executions. Conforming also performs weakly, with a median Accuracy of 0.46 and mean of 0.47, but shows much greater instability, with values ranging from approximately 0.15 to 0.80.

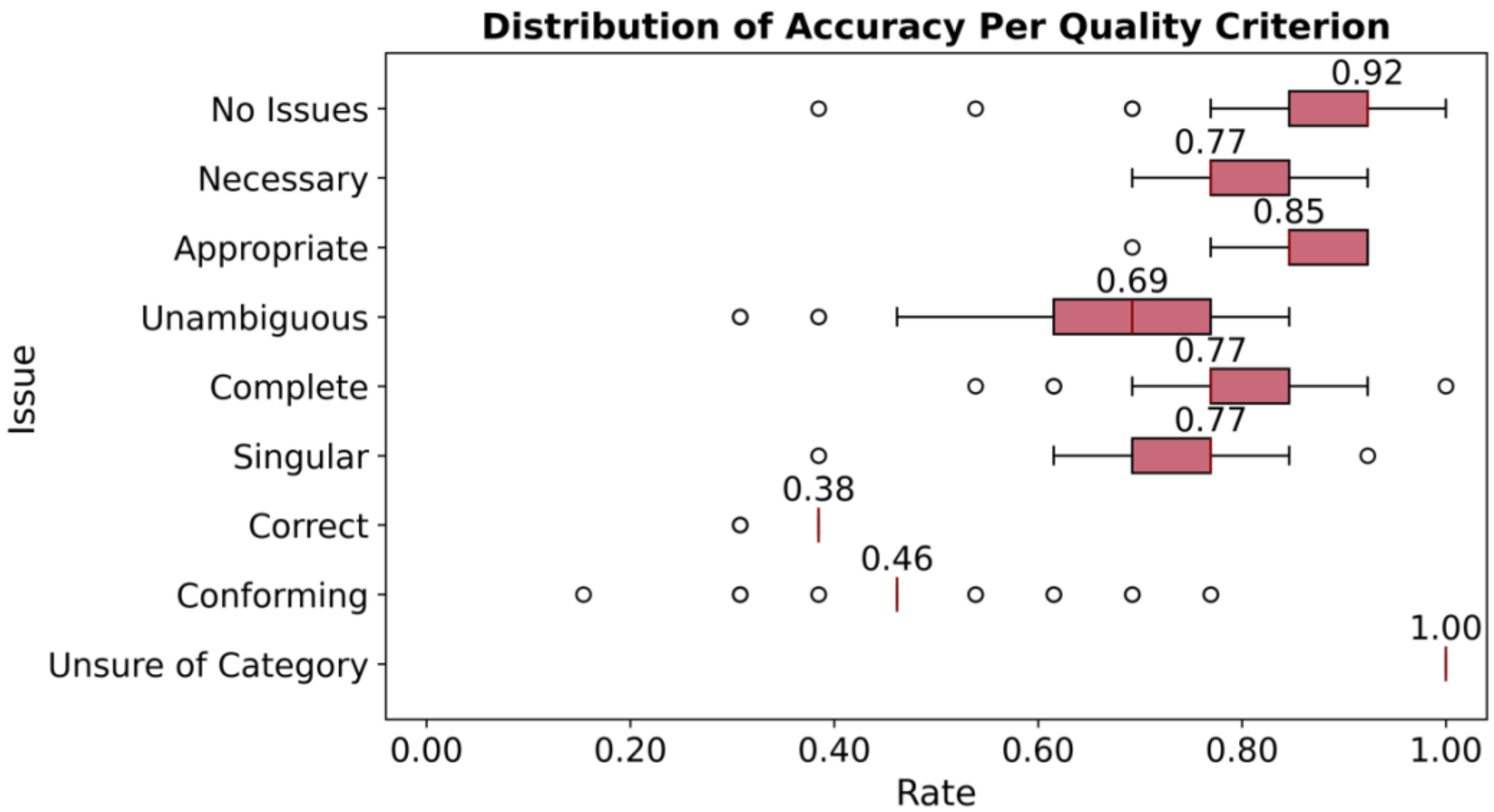

**Figure B3.** Accuracy Distribution by Quality Criterion

Moderate performance is observed for Unambiguous, Singular, Complete, and Necessary. Accuracy for Unambiguous has a median of 0.69 and mean of 0.66, with noticeable variability across executions. Singular, Complete, and Necessary each show stronger central tendencies, with median Accuracy values around 0.77. However, Complete and Singular still exhibit moderate variability.

The strongest performance is observed for Appropriate, No Issues, and Unsure of Category. Accuracy for Appropriate is relatively stable, with a median of 0.85 and mean of 0.88. No Issues shows the strongest non-collapsed performance, with a median of 0.92 and mean of 0.89. Unsure of Category exhibits a collapsed Accuracy distribution at 1.00, indicating perfect and stable classification across all executions. Overall, these results suggest that high aggregate accuracy in some categories is largely driven by reliable non-issue classification, while criteria such as Correct and Conforming remain challenging for the model.

### Appendix B.4. Precision Per Quality Criterion

Figure B4 presents the distribution of Precision across requirement quality criteria, reflecting how reliable the LLM's positive issue predictions are once an issue is flagged. The weakest prediction reliability is observed for No Issues and Necessary, where Precision is collapsed at 0.00. This indicates that positive predictions in these categories are consistently incorrect. Low and unstable reliability is also observed for Singular and Complete. Precision for Singular has a median of 0.29 and mean of 0.21, indicating that singularity-related issue flags are often unreliable. Precision for Complete is slightly higher, with a median of 0.33 and mean of 0.46, but values span nearly the full 0.00 to 1.00 range, showing substantial execution-level instability.

Moderate prediction reliability is observed for Unambiguous, where Precision centers around 0.50 with comparatively stable spread. Stronger reliability is observed for Conforming, with a median Precision of 0.71 and mean of 0.81, indicating that conforming-related issue flags are often valid despite some variability.

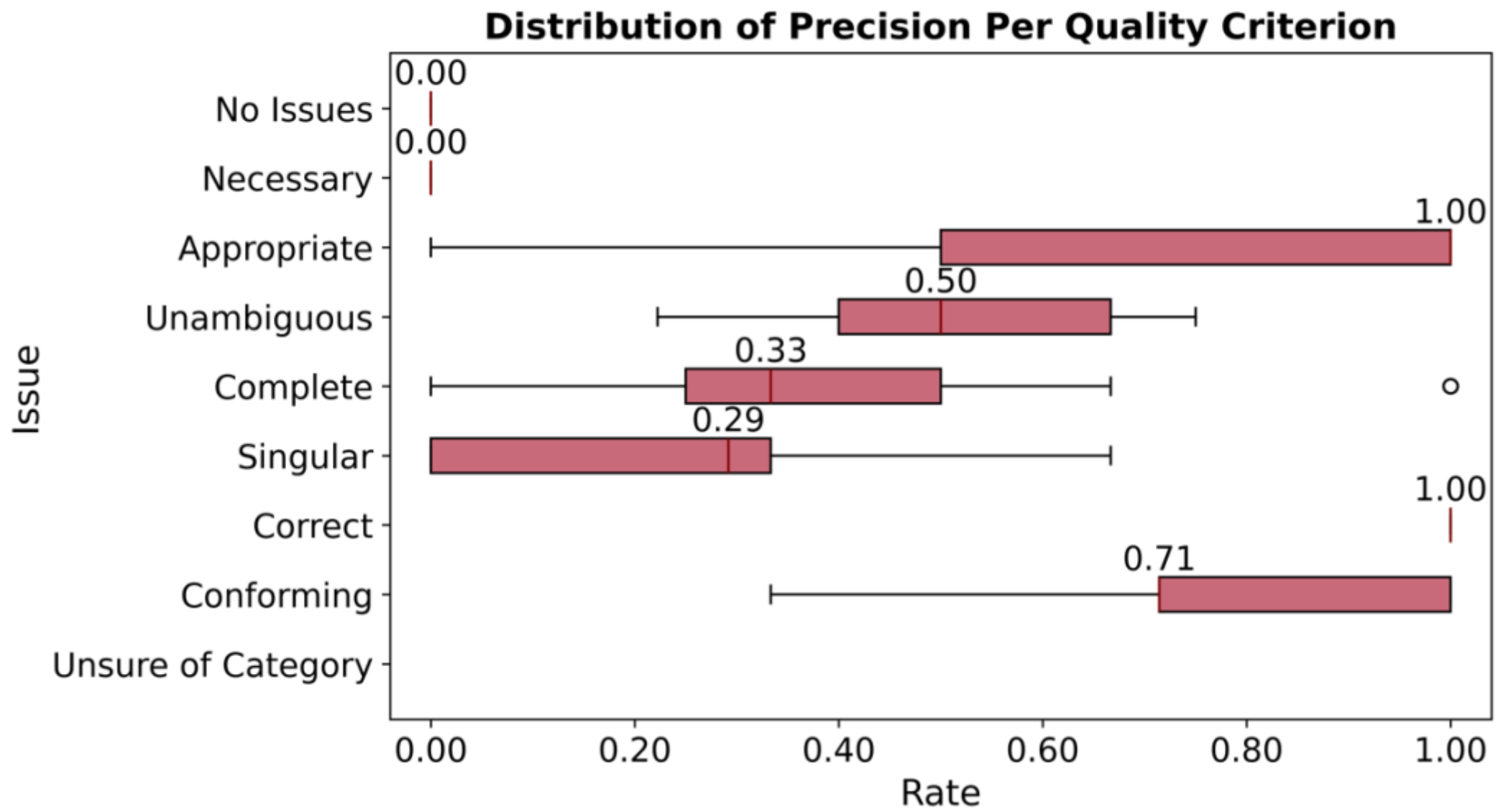

**Figure B4.** Precision Distribution by Quality Criterion

The strongest Precision values are observed for Appropriate and Correct, both with median Precision values of 1.00. However, Appropriate remains unstable, with values ranging from 0.00 to 1.00 and a mean of 0.78. Precision for Correct is collapsed at 1.00, indicating that flagged correctness-related issues are valid when they occur. However, this should not be interpreted as strong overall performance, because the earlier TPR results show that correctness-related issues are rarely detected in the first place. Unsure of Category is omitted because Precision is mathematically undefined for this category.

**Appendix B.5. F1 Per Quality Criterion**

Figure B5 presents the distribution of F1-scores across individual requirement quality criteria, providing a consolidated measure of the LLM's ability to balance issue detection capability (TPR) with prediction reliability (Precision). Among the criteria, the weakest balanced performance is observed for Correct. F1 performance for Correct remains collapsed at 0.20 across all executions, indicating consistently poor balance between detecting correctness-related issues and producing reliable positive predictions. Singular also shows weak balanced performance, with a median F1-score of 0.40 and a mean of 0.44, suggesting that singularity-related issue detection remains limited despite occasional stronger executions.

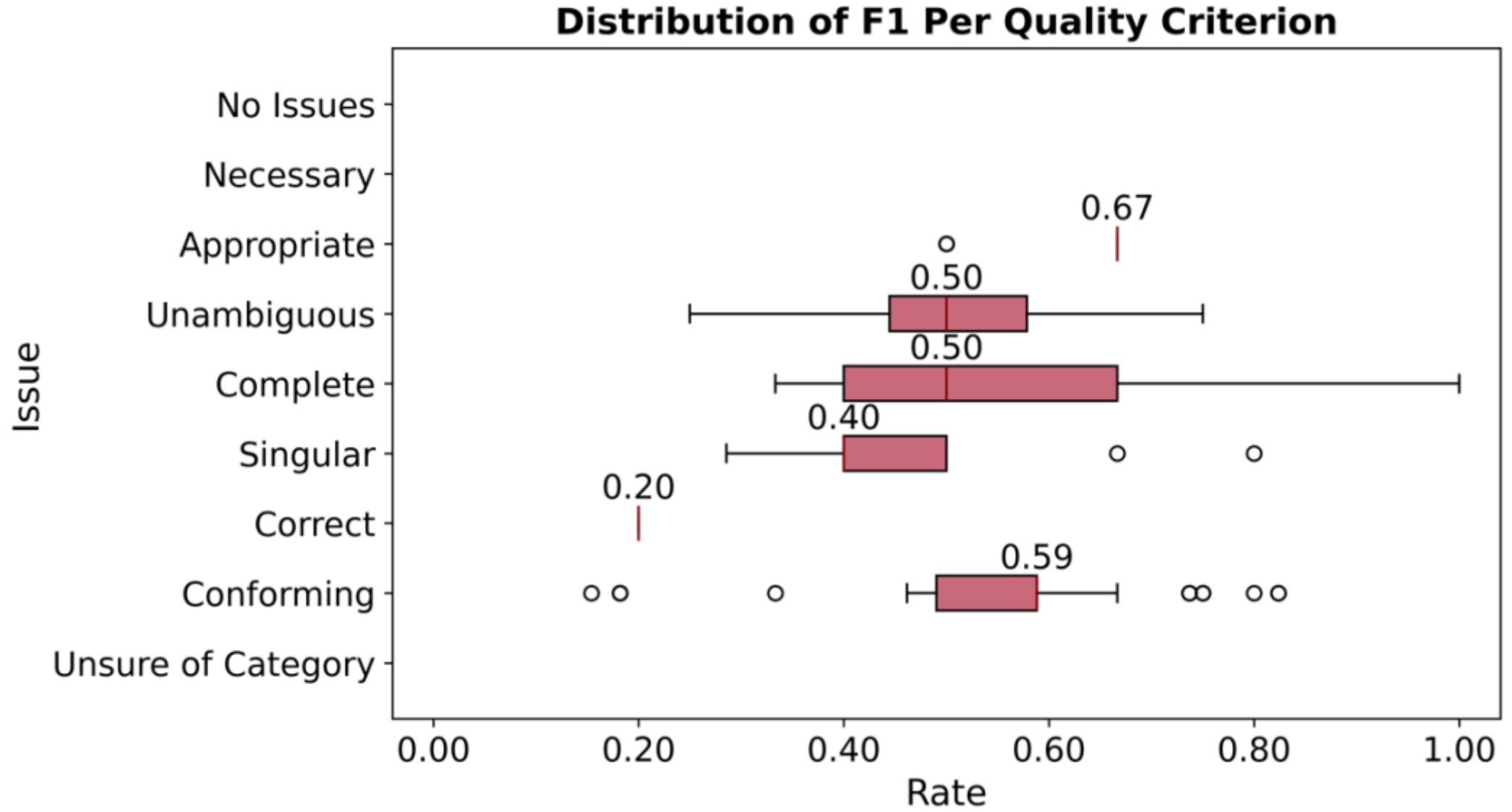

**Figure B5.** F1 Distribution by Quality Criterion

Moderate performance is observed for Unambiguous and Complete. F1 performance for Unambiguous is relatively stable, with a median of 0.50 and mean of 0.52, indicating moderate but consistent balanced performance. Complete shows the same median and mean F1-score, but with greater variability, suggesting that the model's balanced performance for completeness-related issues is less dependable across executions. Stronger performance is observed for Conforming and Appropriate. F1 performance for Conforming has a median of 0.59 and mean of 0.53, indicating comparatively strong but variable balanced performance. The strongest overall F1 performance is observed for Appropriate, with a median of 0.67 and mean of 0.64, suggesting the most consistent balance between issue detection and prediction reliability among the evaluated criteria.

No Issues, Necessary, and Unsure of Category are omitted from Figure B5 because F1 is undefined when the required Precision or TPR components are undefined or when both are zero. Necessary represents the most severe omitted case because the model failed to detect any true positive instances across all executions.